\documentclass[12pt]{article}
\usepackage[T1]{fontenc}
\usepackage{newtxtext,newtxmath}
\usepackage{graphicx}
\usepackage[letterpaper,margin=1in]{geometry}
\renewenvironment{abstract}
	{\quotation}
	{\endquotation}
\date{}

\makeatletter
\renewcommand{\fnum@figure}{\textbf{Figure \thefigure}}
\renewcommand{\fnum@table}{\textbf{Table \thetable}}
\makeatother
\usepackage{scicite}
\usepackage{url}

\usepackage{array}
\usepackage{xcolor}
\def\scititle{
	Unveiling the Importance of Nonshortest Paths in Quantum Networks
}

\title{\bfseries \boldmath \scititle}

\author{
    Xinqi~Hu$^{1}$, Gaogao~Dong$^{1\ast}$, Kim~Christensen$^{2\dagger}$, Hanlin~Sun$^{3}$, Jingfang~Fan$^{4}$, \and
    Zihao~Tian$^{5}$, Jianxi~Gao$^{6,7}$, Shlomo Havlin$^{8}$, Renaud~Lambiotte$^{9,10\ddagger}$, Xiangyi~Meng$^{11\S}$\and
    \small$^{1}$School of Mathematical Sciences, Jiangsu University, Zhenjiang, Jiangsu 212013, China.\and
    \small$^{2}$Blackett Laboratory and Centre for Complexity Science, Imperial College London, London SW7 \small2AZ, UK.\and
    \small$^{3}$Nordita, KTH Royal Institute of Technology and Stockholm University, Hannes Alfvéns väg 12, SE-106 91\and
   \small Stockholm, Sweden.\and
    \small$^{4}$School of Systems Science and Institute of Nonequilibrium Systems, Beijing Normal University, Beijing\and
    \small 100875, China\and
    \small$^{5}$School of Management and Engineering, Nanjing University, Nanjing 210093, China.\and
    \small$^{6}$Network Science and Technology Center, Rensselaer Polytechnic Institute, Troy, New \small York 12180, USA.\and
    \small$^{7}$Department of Computer Science, Rensselaer Polytechnic Institute, Troy, New York 12180, USA.\and
    \small$^{8}$Department of Physics, Bar-Ilan University, Ramat Gan 52900, Israel.\and
    \small$^{9}$Mathematical Institute, University of Oxford, Oxford OX2 6GG, UK.\and
    \small$^{10}$Turing Institute, London NW1 2DB, UK.\and
    \small$^{11}$Department of Physics, Applied Physics, and Astronomy, Rensselaer Polytechnic Institute, Troy, New York\and
    \small 12180, USA.\and
    \small$^\ast$Corresponding authors: Gaogao Dong (dfocus.gao@gmail.com), Kim Christensen (k.christensen@imperial.ac.uk),\and
    \small Renaud Lambiotte (renaud.lambiotte@maths.ox.ac.uk), Xiangyi Meng (xmenggroup@gmail.com)\and
    \small X.H., G.D., K.C., R. L., and X. M. contributed equally to this work.
}

\begin{document} 

\maketitle

\begin{abstract} \bfseries \boldmath
Quantum networks (QNs) exhibit stronger connectivity than predicted by classical percolation, yet the origin of this phenomenon remains unexplored. We apply a statistical physics model---concurrence percolation---to uncover the origin of stronger connectivity on hierarchical scale-free networks, the ($U,V$) flowers. These networks allow full analytical control over path connectivity through two adjustable path-length parameters, $U \leq V$. This precise control enables us to determine critical exponents well beyond current simulation limits, revealing that classical and concurrence percolations, while both satisfying the hyperscaling relation, fall into distinct universality classes. This distinction arises from how they ``superpose'' parallel, nonshortest path contributions into overall connectivity. Concurrence percolation, unlike its classical counterpart, is sensitive to nonshortest paths and shows higher resilience to detours as these paths lengthen. 
This enhanced resilience is also observed in real-world hierarchical, scale-free Internet networks. Our findings highlight a crucial principle for QN design: when nonshortest paths are abundant, they notably enhance QN connectivity beyond what is achievable with classical percolation.
\end{abstract}

\section*{Teaser}
Contrary to classical expectation in network theory, quantum network can be boosted by exploiting nonshortest path connectivity.

\section*{Introduction}
\begin{figure}[t!]
    \centering
	\includegraphics[width=\linewidth]{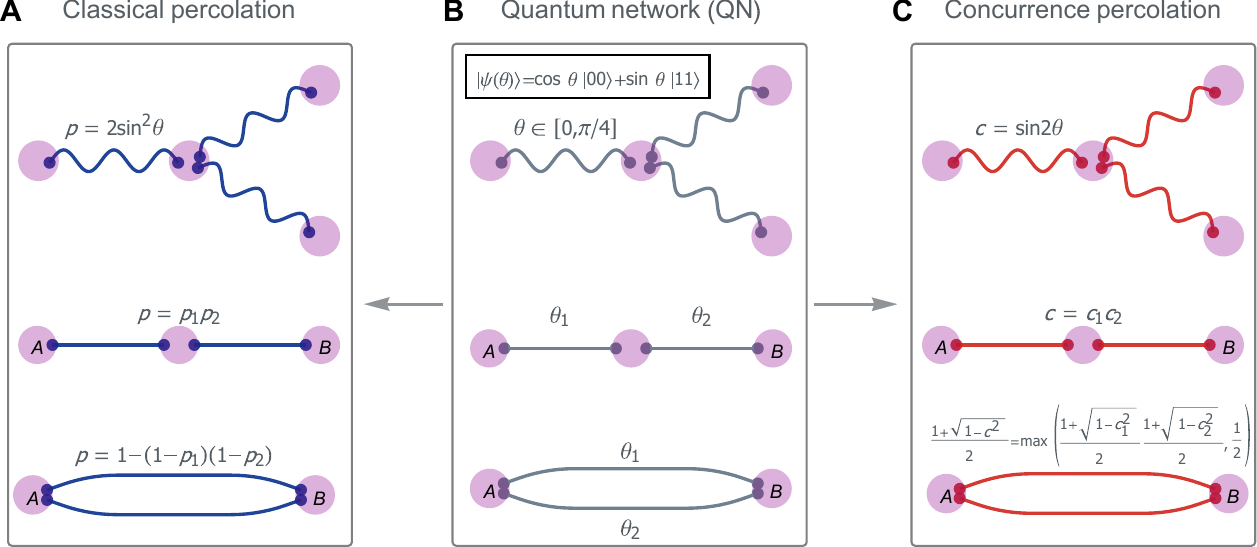}
    \caption{\label{fig1}\textbf{Percolation mappings of a quantum network (QN).}
    (\textbf{A--C})~In a QN, a node comprises several qubits, each entangled with those in other nodes, and a link symbolizes a partially entangled bipartite state $\left|\psi(\theta)\right\rangle$ shared between two nodes. Entanglement transmission across QN can be understood  
    based on two distinct statistical physics mappings: (\textbf{A}) classical percolation (blue) vs.~(\textbf{C}) concurrence ``quantum'' percolation (red). Each mapping is based on its own set of series and parallel path connectivity rules which dictate the ``superposition'' of multiple path connectivities~\cite{bianconi2013statistical} to determine the total percolation connectivity between two distant nodes, say, Alice (node \textit{A}) and Bob (node \textit{B}).
    \hfill\hfill}
\end{figure}

The emerging prospect of {quantum Internet}~\cite{wehner2018quantum} is driving the field of quantum computation and communication to explore larger and more complex scales, entering the realms of statistical physics~\cite{brito2020statistical} and network science~\cite{nokkala2023complex}.
This promising potential of quantum communication relies on the ability of transmitting quantum resources---most notably, entanglement~\cite{Entanglement_Resource}---across vast distances.
Entanglement is fundamental in various quantum information tasks, including distributed quantum computation~\cite{CALEFFI2024110672}, where qubits act as entangled remote proxies via quantum teleportation~\cite{q-teleport_bbcjpw93}, and unbreakable quantum secret sharing~\cite{q-secret-share_g00}, which leverages the no-cloning principle~\cite{q-no-clone_p70}.
However, in practical scenarios, real-world entanglement is often imperfect and susceptible to noise, which poses significant challenges for long-distance transmission~\cite{pirandola2017fundamental}.
To investigate how imperfect entanglement behaves across complex network topologies, a \emph{quantum network} (QN) model~\cite{QEP_acl07} has been introduced.
In the QN model, each link represents an identical, partially entangled state $\left|\psi(\theta)\right\rangle=\cos\theta\left|00\right\rangle+\sin\theta\left|11\right\rangle$, shared between two qubits that are situated at the nodes associated with the link (Fig.~\ref{fig1}). 
The parameter $\theta\in[0,\pi/4]$ serves as a ``link weight'' that quantifies the degree of entanglement. Specifically, $\theta=\pi/4$ signifies a maximally entangled state, while  $\theta=0$ indicates zero entanglement.
This homogeneous, pure-state QN maps onto a {classical percolation} problem~\cite{QEP_acl07}, where each link can employ certain quantum operations as a sort of ``gambling'' to enhance its level of entanglement, allowing a probability $p\equiv 2\sin^2\theta \in [0, 1]$ of becoming perfectly entangled and 
$1-p$ of losing all entanglement~\cite{QEP_acl07}. 
Within this percolation analogy, 
entanglement transmission between two distant nodes---say, Alice (\textit{A}) and Bob (\textit{B})---translates to the probability of having a path comprising only maximally entangled links between \textit{A} and \textit{B}. The presence of such a path further ensures that 
infinite-distance entanglement can be established between \textit{A} and \textit{B}~\cite{conpt_mgh21}, marking a successful event of entanglement transmission.

In the thermodynamic limit (i.e.,~when the number of nodes $N\to \infty$), classical percolation exhibits a {critical threshold} $p_{\text{th}}$~\cite{galam1996universal,cohen2002percolation}. 
It implies that if $\theta <\sin^{-1}{\sqrt{p_{\text{th}}/2}}$, no entanglement transmission can happen between \textit{A} and \textit{B} when their distance tends to infinity.
This is, however, not the case in 
the QN. Here, a more effective mapping from QN to a percolation-like statistical theory of \emph{concurrence percolation} has been found~\cite{conpt_mgh21}.
This mapping, while referred to as percolation, is based on a different framework that moves away from the traditional cluster-based concepts such as cluster-size distribution. Instead, it is grounded in Kesten's original treatment of classical percolation, which focuses on \emph{paths}~\cite{cross-probab-square_k80}. Within its path-based framework,  both classical and concurrence ``quantum'' percolating connectivity  
can be understood as a ``superposition'' of the contributions of all possible paths connecting nodes \textit{A} and \textit{B}.
In classical percolation, the overall probability of connection of \textit{A} and \textit{B} can be computed through each individual path,
contingent upon the parameter $p= 2\sin^2\theta \in [0, 1]$. The superposition can be simplified into connectivity rules, such as the series and parallel rules (Fig.~\ref{fig1}A). Similarly, in concurrence percolation, 
a distinct parameter $c$ denoting \emph{concurrence}, a conventional measure of entanglement, can be defined via $\theta$ in the QN, $c\equiv \sin 2\theta \in [0, 1]$. The superposition of concurrence paths adheres to a distinct set of connectivity rules (Fig.~\ref{fig1}C). Concurrence percolation also reveals a critical threshold $c_{\text{th}}$, which, in terms of $\theta$, is always lower than $p_{\text{th}}$: $\left(\sin^{-1}{c_{\text{th}}}\right)/2\le \sin^{-1}{\sqrt{p_{\text{th}}/2}}$~\cite{conpt_mhtdlgh23}, indicating that QNs have {stronger} connectivity than classical percolation predicts.

A challenging and fundamental  question thus arises. What is the origin of this stronger connectivity, from a {statistical physics} perspective? It is conceivable that concurrence percolation might simply be a variant of classical percolation (albeit under a different set of variables), thereby belonging to the same universality class. Alternatively, concurrence percolation could represent a fundamentally distinct phenomenon from classical percolation, characterized by distinct critical exponents. Previous research, unfortunately, has been unable to identify  the universality class of concurrence percolation due to computational constraints~\cite{ConPT_PRL}. Gaining a deeper insight into the nature of this stronger concurrence connectivity could shed light on QN design from \emph{first principles}, as explored through the lens of network science.

This study examines the critical phenomena of concurrence percolation in networks with hierarchical and scale-free structures~\cite{song2006origins,gallos2007scaling}, which are typical characteristics of many real-worlds networks, including the {Internet}~\cite{tilch2020multilayer}, transportation networks~\cite{wang2020unraveling}, and brain networks~\cite{moretti2013griffiths, sporns2005human}.
Firstly, we exactly determine the critical exponents for a family of hierarchical scale-free network models known as the ($U,V$) flowers~\cite{Fractal_PRE,RHD_PRE2}, which are characterized by two distinct network length scales, $U \leq V$. The ($U, V$) flowers provide a simple model 
for studying the influence of varying length scales through both analytical and numerical methods. Our analysis firmly shows that concurrence ``quantum'' percolation belongs to a universality class distinct from 
that of classical percolation.

More importantly, we highlight that this separation of universality classes is rooted in how the two percolation problems
respond to an increase in the longer length scale, $V$, controlling the lengths of \emph{nonshortest paths}. When $V\to \infty$, we find that the classical percolation critical exponents become decoupled from $V$, depending \emph{only} on the shorter length scale $U$. In contrast, the concurrence percolation critical exponents depend on \emph{both} $U$ and $V$. 
This distinction extends to the behavior of critical thresholds: while both $p_{\text{th}}$ and $c_{\text{th}}$ tend towards unity as $V\to \infty$, the concurrence threshold $c_{\text{th}}$ has a slower rate of approach.
This implies a {higher resilience}~\cite{dong2021optimal,gao2016universal,liu2022network} of concurrence connectivity against increase 
of $V$, a phenomenon we also observe in real-world network topology.
Our findings emphasize the role of {nonshortest paths} in QN: despite the exponential decay of entanglement along longer paths, these paths {cannot} be ignored in concurrence percolation. In fact, \emph{if abundant, nonshortest paths still contribute notably to QN connectivity.
} In practice, this principle suggests that an effective design of QN should move beyond focusing solely on the shortest paths to strategic incorporation of longer paths as well. This view may open up novel opportunities to explore advanced quantum communication technologies, such as path routing~\cite{q-netw-route_p19} and network coding~\cite{5513644}, in a synergistic manner. By highlighting the role of nonshortest paths, this work provides insight into the fundamental mechanism driving the superior performance of QNs and offers guidance for designing robust QNs.

\section*{Results}
\subsection*{Critical Exponents and Hyperscaling} 
\begin{figure}[ht!]
    \centering
    \includegraphics[width=\linewidth]{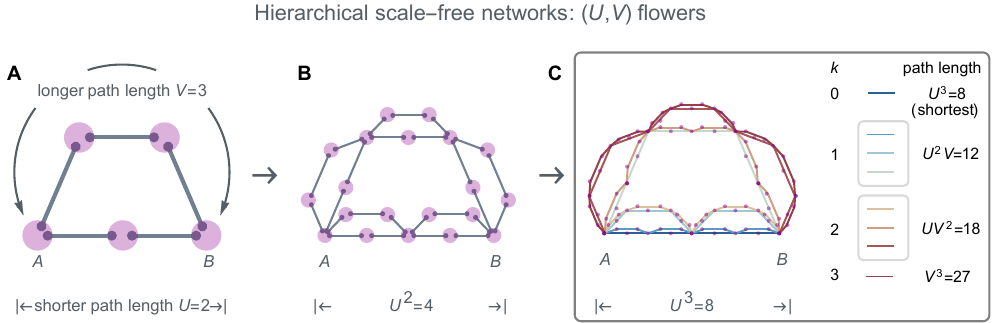}
    \caption{\label{fig_hierarchical net}\textbf{($U,V$) flowers as hierarchical scale-free networks.} 
    The ($U,V$) flower is a self-similar network, constructed by iteratively replacing each link with a motif comprising two nodes connected by two parallel paths with lengths $U \le V$, respectively. The $n$-th generation ($U,V$) flower can be fully decomposed into $2^n$ nonoverlapping paths between nodes \textit{A} and \textit{B}, with varying lengths $\{ U^n, U^{n-1} V, \dots, V^n \}$ and corresponding count of paths $\{C^{0}_{n}, C^{1}_{n}, \dots, C^{n}_{n}\}$, where $C^{k}_{n}=\frac{n!}{k!(n-k)!}$.
    The example shows $U=2$ and $V=3$ for (\textbf{A}) $n=1$, (\textbf{B}) $n=2$, and (\textbf{C}) $n=3$. In (\textbf{C}) the 8 nonoverlapping paths between \textit{A} and \textit{B} are indicated with different colors.}
    \hfill\hfill
\end{figure}

We begin by focusing on ($U,V$) flowers~\cite{Fractal_PRE}, a canonical example of hierarchical, scale-free networks. In a ($U,V$) flower,
the $(n+1)$-th generation is built by replacing each existing link in the $n$-th generation by a basic motif, formed by a shorter path with $U$ links and a longer path with $V$ links between two nodes (Fig.~\ref{fig_hierarchical net}). 
The two scales $U$ and $V$ determine different network characteristics: while the shorter length scale $U$ governs the unique shortest path length,
\begin{equation}
\label{eq_diameter}
    L=U^n,
\end{equation}
between nodes \textit{A} and \textit{B}, which is asymptotically the same as the diameter of the network~\cite{RHD_PRE2}, 
the longer length scale $V$ controls the lengths of all other nonshortest paths, 
thereby controlling the fractal dimension of the network, 
\begin{equation}
d= \lim \limits _{n \to \infty}\frac{\ln N}{\ln L} = \frac{\ln(U+V)}{\ln U},
\end{equation}
where $N\sim \left(U+V\right)^{n}$ is the number of nodes in the network~\cite{RHD_PRE2}.
We find that the ($U,V$) flower is not hyperbolic except in the case of $U=1$ (SI, Section~S1). Hyperbolic networks, known for their distinctive critical phenomena such as Berezinskii--Kosterlitz--Thouless-type transitions and discontinuous percolation behaviors~\cite{boettcher2012ordinary, bianconi2018topological, bianconi2019percolation, kryven2019renormalization,sun2020renormalization}, are beyond the scope of this study. Therefore, we exclude $U=1$ from our analysis and only consider $1< U\le V$.

An interesting feature of ($U,V$) flowers is that they are series-parallel networks~\cite{series-parallel-netw_d65}. Consequently, they allow for an analytical calculation of 
the \emph{sponge-crossing probability} in classical percolation~\cite{cross-probab-square_k80, cross-probab-triangle_w81}, $P_{\text{SC}}$ (a.k.a.~the spanning probability $\Pi_\infty (p;L)$ in boundary conformal field theories~\cite{conformal-percolation_c01}). For ($U,V$) flowers, $P_{\text{SC}}$ denotes the probability of connection between the two boundary nodes \textit{A} and \textit{B}.
Specifically, $P_{\text{SC}}$ can be calculated through iteratively applying an exact renormalization-group (RG) function that is constructed from series and parallel rules (Fig.~\ref{fig1}A):
\begin{eqnarray}
\label{sponge-crossing}
\mathcal{R}(p)=\text{para}(\text{seri}(\stackrel{U}{\overbrace{p,p,\dots,p}}),\text{seri}(\stackrel{V}{\overbrace{p,p,\dots,p}})).
\end{eqnarray}
Equation~\eqref{sponge-crossing} allows us to derive the $n$-th generation $P_{\text{SC}}$ by nesting $\mathcal{R}$ a total of $n$ times, given by $P_{\text{SC}}=\stackrel{n}{\overbrace{\mathcal{R}(\mathcal{R}(\mathcal{R}(\dots \mathcal{R}}}(p))))$.
The critical threshold $p_{\text{th}}$ can be determined by finding the nontrivial fixed point $p^{*}$ that satisfies the RG fixed point equation, $\mathcal{R}(p^{*})=p^{*}$.

\begin{figure}[ht!]
    \centering
    \includegraphics[width=\linewidth]{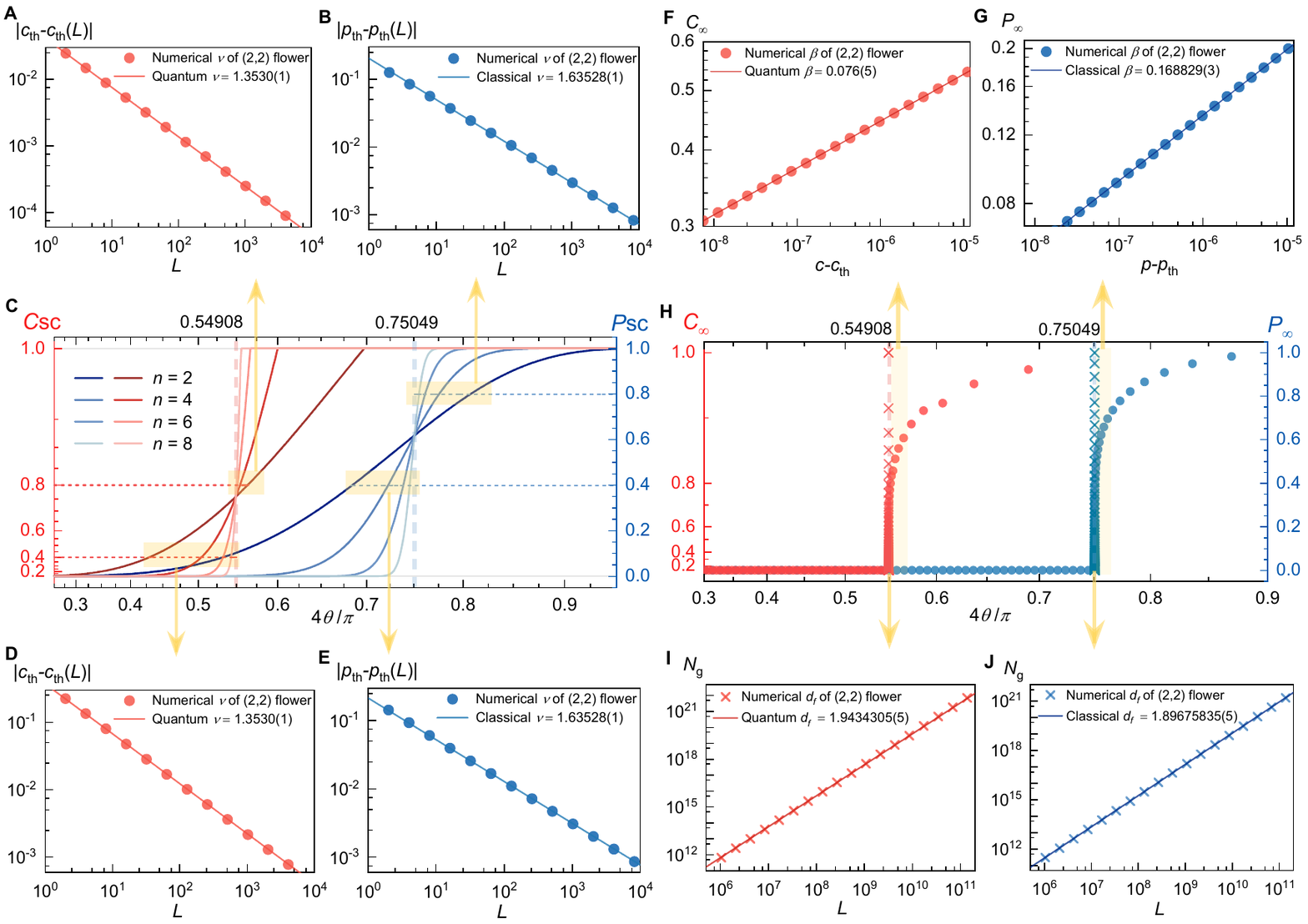}
    \caption{\textbf{Critical phenomena on ($U,V$) flowers.}
    (\textbf{A--B})~Scaling of critical exponent $\nu$ in quantum (red) and classical (blue) mappings above the thresholds, where $c_{\text{th}}(\textit{L})$ and $p_{\text{th}}(\textit{L})$ are solved at $C_{\text{SC}}=0.8$ and $P_{\text{SC}}=0.8$ respectively and $n=1,2,\dots,13$. (\textbf{C})~Classical and concurrence sponge-crossing percolation of generation $n=2,4,6,8$ on ($2,2$) flowers. The dashed vertical lines are at the critical thresholds. 
    (\textbf{D--E})~Scaling of $\nu$ in quantum and classical mappings below the thresholds, where $c_{\text{th}}(\textit{L})$ and $p_{\text{th}}(\textit{L})$ are solved at $C_{\text{SC}}=0.4$ and $P_{\text{SC}}=0.4$ respectively and $n=1,2,\dots,13$. 
    (\textbf{F--G})~Scaling of classical and quantum critical exponent $\beta$. 
    (\textbf{H})~Percolating strengths $P_{\infty}$ and $C_{\infty}$ on ($2,2$) flowers for $n=150$.  
    (\textbf{I--J})~Scaling of ``percolating cluster'' $N_g = NP_{\infty} \propto L^{d_f}$ (classical) and $N_g = NC_{\infty} \propto L^{d_f}$ (quantum) with network diameter $L$. The cross symbols are the percolation value at the critical threshold for $n=20,21,\dots,37$.
    \hfill\hfill}
    \label{fig_percolations}
\end{figure}

Similarly, the sponge-crossing concurrence $C_{\text{SC}}$ and the concurrence threshold $c_{\text{th}}$ (hereafter referred to as ``quantum'' results) can be calculated by replacing $P_{\text{SC}}$ by $C_{\text{SC}}$ in Eq.~\eqref{sponge-crossing} and the corresponding classical series/parallel rules to their quantum counterparts (Fig.~\ref{fig1}A). 
For example, in a ($2,2$) flower, for concurrence percolation, we find $c_{\text{th}} = 0.759\dots$, or $\theta \approx 0.549 \pi/4$, 
while for classical percolation, $p_{\text{th}}=(\sqrt{5}-1)/2 \approx 0.618$~\cite{RHD_PRE2}, corresponding to $\theta \approx 0.750 \pi/4$ (Fig.~\ref{fig_percolations}C). This is an explicit example showing
that the quantum threshold is lower than the classical one~\cite{conpt_mgh21}.
Furthermore, we show that the critical threshold is unique for both classical and quantum percolation (SI, Section~S2), indicative of an ordinary second-order continuous phase transition.

Near criticality, the thermal exponent $\nu$ 
characterizes the divergence of the correlation length as one approaches the critical threshold.
We can obtain an exact expression for $\nu$ by evaluating~\cite{christensen2005complexity} 
\begin{equation}
\label{simulation nu}
\frac{\partial \mathcal{R}(p)}{\partial p}\big|_{p=p_{\text{th}}} =  U ^ {1/\nu} 
\Rightarrow \nu=\frac{\ln{U}}{\ln \left(\frac{\partial \mathcal{R}(p)}{\partial p}\big|_{p=p_{\text{th}}}\right)},
\end{equation}
where $U$ corresponds to the length scale that 
is associated with the network diameter, see Eq.~\eqref{eq_diameter}. 
Compared to the established classical value of $\nu = 1.635\dots$~\cite{RHD_PRE2}, we find a 
quantum value of $\nu = 1.352\dots$. These exact values are numerically verified by finite-size scaling analysis: assuming that the finite-size critical threshold $p_{\text{th}}(L)$ deviates from the true threshold $p_{\text{th}}$ by $|p_{\text{th}}(L)-p_{\text{th}}| \sim L^{-1/\nu}$, we find $\nu \approx 1.63528(1)$ for classical percolation and $\nu \approx 1.3530(1)$ for quantum percolation on both sides of the critical threshold of the ($2,2$) flower (Figs.~\ref{fig_percolations}A,B,D,E).

In the traditional study of classical percolation, not only does the length scale but the cluster size distribution shows critical behaviors as well. Notably, the emergence of a unique infinite percolating cluster when $p >  p_{\text{th}}$ leads to a scaling behavior $P_\infty \sim (p-p_{\text{th}})^ {\beta}$ for $p \to p_{\text{th}}^{+}$, where the percolating strength $P_{\infty}=N_g/N$ represents the percolating cluster's relative size $N_g$ to the total network size $N$, that is, the probability that a randomly chosen node belongs to the percolating cluster~\cite{christensen2005complexity}.
This cluster-based concept of $P_{\infty}$ has not been translated into the quantum counterpart, which relies on path connectivity rules and lacks a notion of ``clusters.'' 
Nevertheless, we notice that $P_{\infty}$ can be alternatively considered as \emph{the probability of a randomly chosen node in the bulk to reach the boundaries of the network}~\cite{RHD_PRE2}.
This consideration allows us to redefine $P_\infty$, and crucially $C_\infty$, in terms of path connectivity from a randomly chosen node (which must be averaged across all nodes) to the boundaries (which are assumed to exist). For the ($U,V$) flowers, the boundaries are represented by \textit{A} and \textit{B}. The construction is detailed in SI, Sections~S3 and S4.

Using this alternative definition of $P_\infty$, we find an exponent $\beta\approx 0.168829(3)$ for classical percolation (Fig.~\ref{fig_percolations}G), closely matching the exact cluster-defined value, $\beta=0.165\dots$ on ($2,2$) flowers~\cite{RHD_PRE2}.
The numerical results does not agree exactly with the theoretical value because computing $P_{\infty}$ requires higher-order path connectivity rules that extend beyond the basic series and parallel rules. We approximate these higher-order rules using only the series/parallel rules through a technique known as the star-mesh transform~\cite{conpt_mgh21}, which yields results that are close, but not exact (SI, Section~S6). In the quantum domain, we observe $\beta \approx 0.076(5)$ (Fig.~\ref{fig_percolations}F), markedly distinct and smaller than 
its classical counterpart, suggesting an almost discontinuous transition that is comparable to explosive percolation [$\beta\approx 0.0555(1)$]~\cite{chen2013phase,explos-percolation_dsggna19}.

Another critical exponent, $d_f$, defines the fractal dimension of the percolating cluster at the critical threshold, with $N_g \sim L^{d_f}$. Again, lacking a ``cluster'' notion, $N_g=N P_\infty$ must be redefined through $P_\infty$ at the critical threshold. We find $d_{f} \approx 1.89675835(5)$ (Fig.~\ref{fig_percolations}J) in the classical scenario, very close to the cluster-defined value $d_f = 1.899\dots$~\cite{RHD_PRE2}. In contrast, we find that the quantum scenario shows $d_{f} \approx 1.9434305(5)$ (Fig.~\ref{fig_percolations}I).

Interestingly, the three path-defined critical exponents still adhere to the hyperscaling relation, $d_f=d-\beta/\nu$~\cite{book1}, in both classical and concurrence percolation (SI, Section~S5), implying a consistent path-connectivity framework across different percolation theories. As an example, a summary of the critical exponents for the ($2,2$) flower is shown in Table~\ref{table_exponents}. Critical exponents of a general ($U,V$) flower can be obtained via the same approach. See details in Table~\ref{table_asymptotic} for results in the $V \to \infty$ limit and SI, Section~S6.

\begin{table}[h]
	\centering
        \vspace{0cm}
	\caption{\textbf{Numerical critical exponents for $U=V=2$.}
 $d$: network dimension;
    $\nu$: characterizing divergence of the correlation length when $p \to p_{\text{th}}^{\pm}$ ($c \to c_{\text{th}}^{\pm}$); 
    $d_f$: 
    scaling of the percolating cluster $N_g=N P_\infty$ ($N C_\infty$) with system size $N$ at $p=p_\text{th}$ ($c=c_\text{th}$);
    $\beta$: characterizing symmetry breaking of the percolating strength $P_\infty$ ($C_\infty$) when $p \to p_{\text{th}}^{+}$ ($c \to c_{\text{th}}^{+}$); 
    $d-d_f-\beta/\nu$: equal to zero when 
    the hyperscaling relation 
    is satisfied. 
    Numbers in brackets represent standard errors.}
    \vspace{0.5cm}
    \label{table_exponents}
        \setlength{\tabcolsep}{0.5cm}
	\begin{tabular}{c|c|c}
		\hline\hline
		\makebox[0.5cm] & \makebox[3cm]{Classical percolation}& \makebox[6cm]{Concurrence ``quantum'' percolation} \\
		\hline
        $d$ & 2 & 2\\
        \hline
		$\nu$ & 1.63528(1) & 1.3530(1)\\
		\hline
		$d_f$ & 1.89675835(5) & 1.9434305(5)\\
            \hline
		$\beta$ & 0.168829(3) & 0.076(5)\\
		\hline
            $d-d_f-\beta / \nu$ & $3 \times 10^{-9}$ $(2 \times 10^{-6})$ & $4 \times 10^{-4}$ $(4 \times 10^{-3})$ \\
            \hline\hline
	\end{tabular}
        \vspace{0cm}
\end{table}

Note that all length scales are established in the chemical-distance
(``time'') space~\cite{Fractal_PRE}, thus differing from traditional Euclidean length scales $\xi$ by an additional exponent $z$, related by $L\sim \xi^{z}$. In other words, our values of $d$, $d_f$, and $\nu$ correspond to $d/z$, $d_f/z$, and $z\nu$ in the Euclidean scale~\cite{book2}.

\subsection*{Asymptotic dependence on the longer length scale}
To investigate the impact of the longer length scale $V$ and the shorter length scale $U$ on the two percolation theories, we derive the asymptotic behaviors of both critical thresholds $p_{\text{th}}$ and $c_{\text{th}}$ in the $V\to \infty$ limit using constant $U>1$ and observe the simulation values of percolation critical exponent for finite $U$ when $V$ increases (SI, Section~S6), finding that for classical percolation, 
\begin{equation}
\label{eq_pth_infty}
   p_{\text{th}}\simeq 1-\left(\ln\frac{U}{U-1}\right)V^{-1}+O(V^{-2}),
\end{equation}
and, 
for concurrence percolation,
\begin{equation}
\label{eq_cth_infty}
   c_{\text{th}}\simeq 1-\left(\frac{1}{4}\ln V \right)V^{-1}+O(V^{-1}\ln\ln V).
\end{equation}
The origin of the $V^{-1}$ term is due the fact that as $V\to \infty$, loops within the ($U,V$) flower extend to infinite lengths, and the network essentially becomes tree-like~\cite{sandvik2010ground}. This suggests that nodes \textit{A} and \textit{B} are connected by only a single path. As a result, both thresholds approach $1$ when $V\to \infty$. 
However, for classical percolation, the prefactor of $V^{-1}$ depends solely on $U$, signaling the significant role the \emph{shortest paths} play in classical percolation near the critical threshold. In contrast, concurrence percolation shows a unique behavior. The presence of a logarithmic prefactor $\left(1/4\right) \ln V$ highlights the importance of the \emph{nonshortest paths} in concurrence percolation. 
Moreover, this term is fully decoupled from $U$, suggesting that the shorter length scale no longer plays a determining role in $c_{\text{th}}$. For different $U=2,5,10$, on the one hand, the simulation results show the same conclusion as above. On the other hand, the analysis produces different constant limits for the classical case and different speed of going to zero for quantum case, exhibiting the necessity of the existence of shortest path $U$ (SI, Section~S6).

Our finding can also be illustrated as follows. 
At the percolation thresholds, the sponge-crossing connectivity along the longer path is given by
$p_{\text{th}}^V \simeq \left(U-1\right)/U$ and
$c_{\text{th}}^V \simeq  V^{-1/4}$,
respectively when $V \to \infty$ (SI, Section~S6).
In other words, for concurrence percolation, longer path connectivity tends to drop to zero ($c_{\text{th}}^V\to0$) at the edge of dismantlement ($c= c_{\text{th}}$), indicating that longer path connectivity must be exhausted as the network disintegrates in terms of concurrence.
This contrasts with classical percolation, where the longer path connectivity is a nonzero constant ($p_{\text{th}}^V\nrightarrow 0$) at the edge of dismantlement ($p= p_{\text{th}}$), meaning that the network can dismantle \emph{before} longer paths connectivity are exhausted.
Hence, the nonshortest paths play no role in the critical behaviour of classical percolation.

The asymptotic behaviors of the critical exponents are also different. We analytically determine $\nu$ and $d_f$ directly from the solution of $P_{\text{SC}}$ and $P_\infty$ (cf.~$C_{\text{SC}}$ and $C_\infty$), and $\beta$ from the hyperscaling relation $d_f=d-\beta/\nu$. The results are summarized in Table~\ref{table_asymptotic} in the limit $V \to \infty$. We find that in classical percolation, all dominant terms governing $\nu$, $d-d_f$, and $\beta$ are determined by $U$ only, once again signaling the dominance of the shorter paths. In contrast, in concurrence percolation, we notice the emergence of a very slow correction depending on the nonshortest paths ($\sim \ln \ln V$) affecting the dominant terms in $\nu$ and $d_f$ (SI, Section~S6). Interestingly, the corrections counterbalance each other in the calculation of $\beta$, resulting in a constant $\beta=1$ that is independent on both $U$ and $V$. Unexpectedly, this value coincides with the mean-field value of $\beta$ for classical percolation~\cite{book2}. Since $\beta$ is intrinsically tied to the definition of an order parameter, we hypothesize that this unique, constant value of $\beta$ reveals an entirely distinct symmetry associated with the order parameter in concurrence percolation on ($U,V$) flowers.

\begin{table}[ht!]
        \renewcommand\arraystretch{1.5}
	\centering
	\caption{\textbf{Critical exponents in the $V\hspace{-1.5mm}\to\hspace{-0.5mm}\infty$ limit.} Analytical and numerical asymptotic analyses are provided in SI, Section~S6.}
    \label{table_asymptotic}
    \vspace{0.5cm}
    \begin{tabular}{c|l|l}
		\hline\hline
	    & Classical [$\sim \!f(U)+O(V^{-1})$] & Quantum  [$\sim \!g(U, V$)]\\
		\hline
		$\nu$ & 
        $ \frac{\ln U}{\ln\left[1+\left(U-1\right)\ln \frac{U}{U-1}\right]}\!+\!O(\!V^{-1}\!)$ & 
        $ \frac{\ln U }{\ln \ln V}\!+\!O\left(\frac{\ln \ln \ln V}{\left(\ln \ln V\right)^2}\right)$ \\
		\hline
		$d\!-\!d_f\!$ & $ \frac{\!\ln{U}\!-\!\ln{\left[\left(1\!-\!U\right) + 2/\left(\ln\!{\frac{U}{U-1}}\right)\right]}}{\ln{U}} \!+\!O(\!V^{-1}\!)$ 
        & $\frac{\ln \ln V}{\ln U} \!+\!O\left(\frac{\ln \ln V}{\ln V}\right) $ \\
        \hline
		$\beta$ & $       
  \frac{\!\ln{U}\!-\!\ln{\left[\left(1\!-\!U\right) + 2/\left(\ln\!{\frac{U}{U-1}}\right)\right]}} {\ln \left[1+\left(U-1\right)\ln\frac{U}{U-1}\right]}\!+\!O(\!V^{-1}\!)$ &  
            $ 1 \!+\!O\left(\frac{\ln \ln \ln V}{\ln \ln V}\right)$\\
		\hline\hline
	\end{tabular}
\end{table}

\subsection*{Superposition of paths in real-world networks}
To demonstrate the generality of the results obtained on ($U,V$) flowers, we investigate the role of shortest and nonshortest paths in classical and quantum percolation in real networks. Specifically, we employ path decomposition to systematically assess the contributions of shortest and nonshortest paths to overall network connectivity.
It is worth noting that an $n$-th generation ($U,V$) flower can be fully decomposed into $2^n$ \emph{nonoverlapping but intersecting} paths between two boundary nodes \textit{A} and \textit{B} (Fig.~\ref{fig_hierarchical net}C).
In the language of network theory, these paths are edge-disjoint, but not vertex-disjoint. 
Specifically, each of these paths can have a length 
that is given by one of the options $\{U^n, U^{n-1} V, \dots, V^n\}$. 
The number of paths corresponding to length $U^{n-k} V^{k}$ is determined by the binomial coefficient $C^k_{n} \equiv \frac{n!}{k!(n-k)!}$. 

The exact decomposition goes as follows. 
Initially, in the $1$-st generation, there are only two nonoverlapping paths between \textit{A} and \textit{B}, characterized by lengths $U$ and $V$. Progressing to the $2$-nd generation, each link from the previous generation is systematically replaced by the basic motif, necessitating a decision for each link within a previous path: to replace it by either a $U$-length path or a $V$-length path. Consequently, there are two nonoverlapping options: replacing all links entirely by $U$-length paths, or entirely by $V$-length paths. This bifurcation results in each earlier generation's path splitting into two distinct paths. Thus, as a shorthand, each path can be encoded as a string of $n$ characters (e.g.,~``UVVUUV...''), each designated as `U' or `V'. This string reflects the sequence of choices made at each generation, with `U' for choosing $U$-length paths and `V' for $V$-length paths. Each nonoverlapping path can be uniquely identified by such a string. The corresponding path length is $U^{n-k}V^{k}$, with $k=0,1,\dots,n$, representing the count of `V's in the string. For each possible length, there are $C^{k}_{n}$ such choices, which scale as $C^{k}_{n}\simeq \left(2 \pi k\right)^{-1/2} e^{k} k^{-k} n^k$ for $k\ll n$, $k\to \infty$, suggesting that nonshortest paths are exponentially more abundant.

\begin{figure}[t!]
    \centering
    \includegraphics[width=0.6\linewidth]{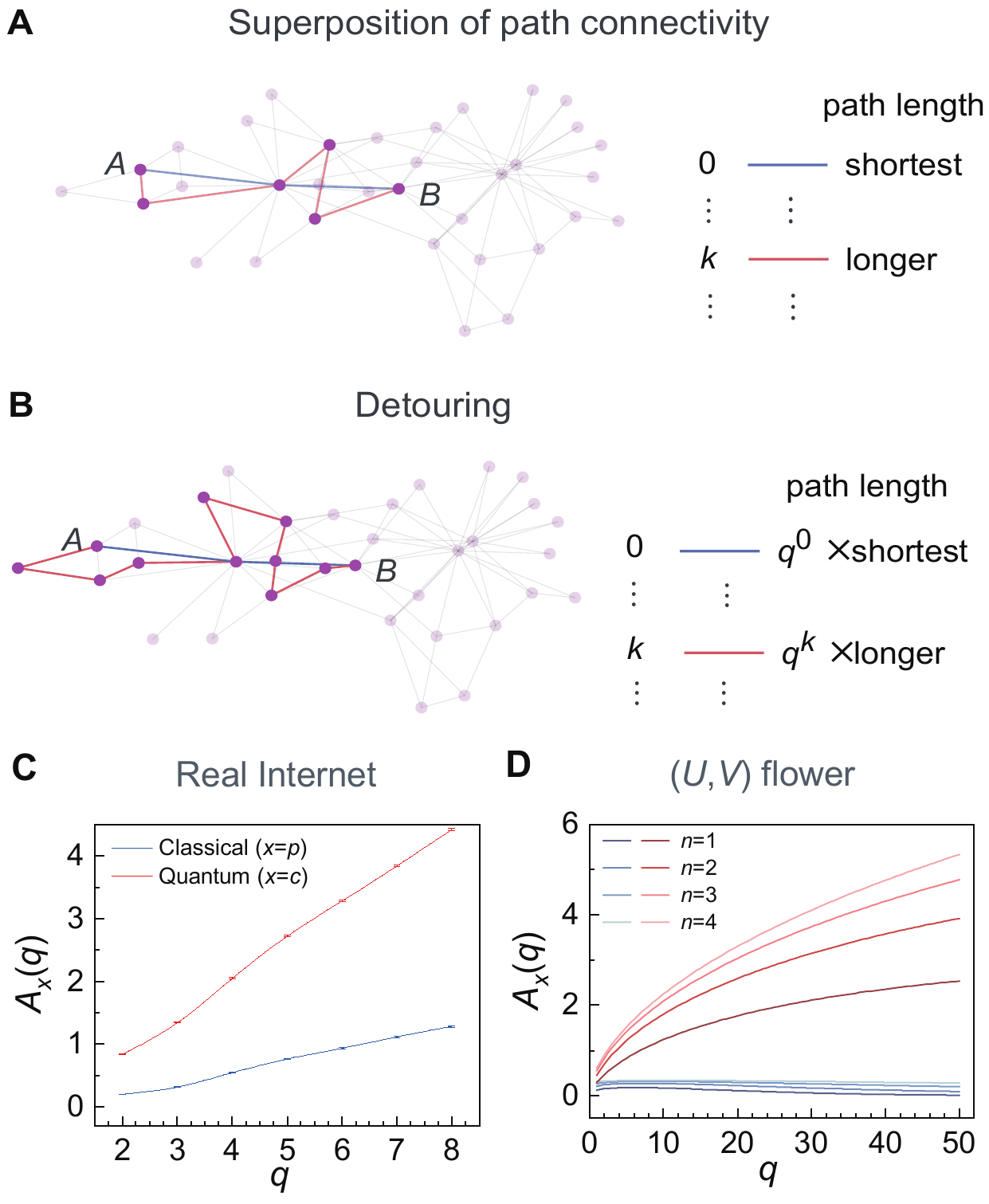}
    \caption{\textbf{General path connectivity and detouring.} (\textbf{A})~Superposition of paths in general networks. Nonoverlapping but potentially intersecting paths are selected between two hub nodes \textit{A} and \textit{B} and grouped by their lengths.
    (\textbf{B})~Detouring of longer paths in (\textbf{A}). Each path in group $k$ is rerouted becoming $q^k$ times longer. The shortest path (blue line) is constant, while the red line shows an increased longer path when $q=2$. (\textbf{C})~The anomalous resilience factors $A_p(q)$ and $A_c(q)$ show different scaling behaviors with respect to the detouring factor $q$ in the Internet network topology~\cite{tilch2020multilayer}. Higher $A_x(q)$ shows higher resilience to detouring. (\textbf{D}) Corresponding patterns are also observed in ($U,V$) flowers. As $q$ increases, the classical $A_p(q)\sim q^{0}$ (blue) approaches a constant value. In contrast, the quantum $A_c(q)\sim \sqrt{q\ln q/8}$ (red) increases with $q$.\hfill\hfill}
    \label{fig_real net}
\end{figure}

The path decomposition is applicable to real networks with more general topologies. 
Specifically, given a real-world network, we select two nodes \textit{A} and \textit{B} and the nonoverlapping (but potentially intersecting) paths between them. These paths are ranked by their lengths and sequentially placed into different groups, labeled by $k=0,1,2,\dots$, such that each successive group encompasses an exponentially larger number of longer paths (Fig.~\ref{fig_real net}A).

We now consider the impact of \emph{detouring} in the real-world network due to damages or congestion. Given that longer paths are more likely to be rerouted, we assume that the paths in group $k$ increase their lengths by a factor of $q^k$, $q\ge 1$ due to detouring. Meanwhile, the number of paths remains constant (Fig.~\ref{fig_real net}B).
To investigate the network's resilience to detouring given the parameter $q$, 
we rewrite both $p_{\text{th}}$ and $c_{\text{th}}$ in terms of $\theta_{\text{th}}^{(p)}$ and $\theta_{\text{th}}^{(c)}$, expressed as
\begin{equation}
\label{Axq}
\theta_{\text{th}}^{(x)}(q)= \theta_{\text{th}}^{(x)}(\infty)-A_x(q) q^{-1},
\end{equation}
where $q^{-1}$ is a common factor, as we expect $A_x(q) q^{-1}\to 0$ when $q\to \infty$ for both classical ($x=p$) and concurrence ($x=c$) percolation. We thus define $A_x(q)\equiv q\left[\theta_{\text{th}}^{(x)}(\infty)-\theta_{\text{th}}^{(x)}(q)\right]$ as the \emph{anomalous resilience factor}, measuring the resilience of $\theta_{\text{th}}^{(x)}(q)$ against approaching the trivial threshold, $\theta_{\text{th}}^{(x)}(\infty)$, which relies solely on the shortest path.

While large-scale QNs are urgently needed, they are still in early stages, hindered by technological challenges in creating quantum memories~\cite{q-mem_saabgdghjkmnprdrrssstwwwwy10,q-netw-mem_mpnk24}. We here focus on existing Internet networks, assuming QNs can follow the same topology. 
Recent experiments transmitting entanglement alongside classical signals via Internet fibers confirm this potential~\cite{q-netw-photon_cekklppstvxkkss22,q-netw_tkk23}.
Here, we analyze a segment of the real-world Internet at the autonomous-system level~\cite{tilch2020multilayer}, representing data forwarding relationships among  administrative entities or domains {(SI, Section~S7)}. 
Given two hub nodes and their nonoverlapping paths, detouring requires replacing each path in group $k$ with a rerouted path from the same Internet that is $q^k$ longer.
We determined $A_x(q)$ for $x=p,c$ for various $q$ (Fig.~\ref{fig_real net}C).

For comparison, we also considered the ($U,V$) flower (Fig.~\ref{fig_real net}D), where rerouting gives rise to new path lengths of $\{U^n, q U^{n-1} V, \dots, q^n V^n\}=\{U^n, \left(q V\right) U^{n-1}, \dots, \left(q V\right)^n\}$, suggesting that $V$ simply adjusts to $q V$. Therefore, rearranging 
Eqs.~\eqref{eq_pth_infty} and~\eqref{eq_cth_infty} leads to the theoretical predictions
$A_p(q)\simeq \ln{\left[U/\left(U-1\right)\right]}/2=O(q^0)$, versus $A_c(q)\simeq V^{1/2} \sqrt{q\ln q/8}$.
The similarity of the dependencies of $A_x(q)$ on $q$ between the Internet and the ($U,V$) flower highlights a shared quantum characteristic: the faster increase of $A_x(q)$ for $x=c$, which indicates a higher resilience of concurrence connectivity. Our observation indicates the presence of such resilience in general network topologies, characterized by nonoverlapping but intersecting paths. 
The distinct universality classes of classical and quantum percolation on ($U,V$) flowers, driven by their differing dependencies on nonshortest paths, suggest that these findings are broadly applicable to other network types, including real-world networks with complex connectivity patterns. 

Note that given the very slow corrections $\sim \ln\ln V$ in the quantum critical exponents (Table~\ref{table_asymptotic}), this extreme effect becomes observable only for a very large length scale, exceeding the Internet diameter.
However, if we consider ``fundamental''  networks, such as spin networks in loop quantum gravity~\cite{loop-q-gravity_ab21}, where the length unit is extremely small (the Planck scale $10^{-35}$~m), the corrections could potentially become detectable.

\section*{Discussion}

In conclusion, our findings show the crucial role of nonshortest quantum entanglement paths in enhancing connectivity within quantum networks (QN). This advancement builds on the theoretical breakthroughs achieved in previous studies of hierarchical scale-free network models~\cite{Fractal_PRE,RHD_PRE2}. Unlike traditional lattice models, these innovative models allow for precise analytical analysis of the complex critical phenomena underlying quantum networks, marking a substantial step forward in the field. 

From the theoretical, statistical physics perspective, our findings, which place concurrence percolation in a distinct universality class to classical percolation, rule out the possibility that differences between classical and concurrence percolation theories are confined merely to short-range (i.e.,~ultraviolet) details. Instead, variations stem from long-range (i.e.,~infrared) behaviors. 
This distinction emphasizes the fact that concurrence percolation represents a fundamentally different statistical framework compared to its classical counterpart. The question of whether concurrence percolation can be described by a field theory akin to the $\phi^3$ model~\cite{potts_a76} for classical percolation remains open. 
Such a ``concurrence'' field theory could simultaneously manifest two (or more) length scales, a common phenomenon in quantum critical systems~\cite{shao2016quantum}.
To this end, a crucial step involves defining a cluster-based order parameter (field) specific to concurrence percolation. Our noncluster-based definition of the percolating strength $C_{\infty}$ on ($U,V$) flowers could potentially serve as a foundation for constructing a cluster-based order parameter. 

From the perspective of quantum computation and communication, our research indicates that it is possible to achieve entanglement transmission via longer, exponentially weaker paths composed of partially entangled links, provided there is a compensatory increase in the number of such paths. This principle finds a parallel in branching processes, where infinite connectivity is achievable if the growing number of trees with given size exactly compensates for the exponential decline in the probability of such a tree existing at the critical threshold~\cite{christensen2005complexity}. Further exploration of this concept through other models, such as the ($U,V,W$) flowers (with three different length scales), could provide deeper insights on the entanglement distribution efficiency in QNs.
The principle of leveraging multiple nonshortest paths is particularly relevant to the rapid advancement of multiplexing techniques in quantum communication~\cite{q-netw-route_lphnm20}. 
However, the unavoidable mixing of entangled pure states with classical noise complexifies the scenario, underscoring the need for a hybrid model that incorporates elements of both classical and concurrence percolation.
We expect that a stronger connectivity would still exist in mixed-state QNs, which, if feasible for use, must go beyond any completely classical treatments.

\clearpage

\bibliography{Longerpath}
\bibliographystyle{sciencemag}

\section*{Acknowledgments}
We thank Jürgen Kurths, Yu Tian, Ginestra Bianconi, Zhuan Li, and Yanxuan Shao for useful discussions.
\paragraph*{Funding:}
G.D. is supported by grants from the National Natural Science Foundation of China (Grant Nos. 62373169, 72174091), the National Key Research and Development Program of the Ministry of Science and Technology of China (Grant No. 2020YFA0608601), and the Young Academic Leaders of the ``Blue Project'' in Jiangsu Province, and the Special Science and Technology Innovation Program for Carbon Peak and Carbon Neutralization of Jiangsu Province (grant no. BE2022612-4). 
R.L. acknowledges support from the EPSRC Grants EP/V013068/1 and EP/V03474X/1, as well from the International Exchanges IEC\textbackslash NSFC\textbackslash 201180 of the Royal Society. 
J.G. acknowledges the support of the US National Science Foundation under Grant No.~2047488, and the Rensselaer---IBM AI Research Collaboration.
S.H.~thanks the Israel Science Foundation (Grant No.~189/19), the EU~H2020 DIT4Tram (Grant number~953783) and the Horizon Europe grant OMINO (Grant number~101086321). 
J.F. thanks the National Natural Science Foundation of China (grant nos. 42450183, 12275020, 12135003, 12205025, 42461144209), the Ministry of Science and Technology of China (2023YFE0109000) and the Fundamental Research Funds for the Central Universities
H.S. acknowledges the support of the Wallenberg Initiative on Networks and Quantum Information (WINQ).
Z.T. is supported by the National Natural Science Foundation of China (grant no. 72401127).
\paragraph*{Author contributions:}
X.M., S.H., and  G.D. conceived the project. 
X.H., X.M., K.C., G.D., H.S, and J.G. designed the methods, validated the data, and performed the analysis. 
X.H., G.D., X.M., R.L., and K.C. performed software development/implementation. 
X.H., G.D., X.M, K.C, and Z.T. curated the data. 
X.H., X.M., G.D., and K.C. performed data visualization. 
X.H., X.M., R.L.,  and G.D. wrote the original draft. 
R.L., X.M., G.D., K.C., X.H., J.G., H.S., and J.F. wrote the revisions. 
G.D., R.L., X.M., and S.H. supervised the project and performed project administration. 
R.L., G.D., J.G., S.H., J.F., H.S., and Z.T. acquired funding.
\paragraph*{Competing interests:}
The authors declare they have no competing interests.
\paragraph*{Data and materials availability:}
All data needed to evaluate the conclusions in the paper are present in the paper and/or the Supplementary Materials. Data of the real AS internet network are also available at https://zenodo.org/record/1038572.

\newpage

\renewcommand{\thefigure}{S\arabic{figure}}
\renewcommand{\thetable}{S\arabic{table}}
\renewcommand{\theequation}{S\arabic{equation}}
\renewcommand{\thepage}{S\arabic{page}}
\setcounter{figure}{0}
\setcounter{table}{0}
\setcounter{equation}{0}
\setcounter{page}{1} 

\begin{center}
\section*{Supplementary Materials for\\ \scititle}

Xinqi~Hu,
Gaogao~Dong$^{\ast}$,
Kim~Christensen$^\dagger$,
Hanlin~Sun,
Jingfang~Fan,
Zihao~Tian,
Jianxi~Gao,
Shlomo~Havlin,
Renaud~Lambiotte$^\ddagger$,
Xiangyi~Meng$^\S$\\ 
\small$^\ast$Corresponding author. Email: Gaogao Dong (dfocus.gao@gmail.com), Kim Christensen (k.christensen@imperial.ac.uk), Renaud Lambiotte (renaud.lambiotte@maths.ox.ac.uk), Xiangyi Meng (xmenggroup@gmail.com)\and

\end{center}

\subsubsection*{This PDF file includes:}
Sections S1 to S7\\
Figures S1 to S9

\newpage
\tableofcontents

\renewcommand{\thesection}{S\arabic{section}}
\renewcommand{\thefigure}{S\arabic{figure}}
\renewcommand{\theequation}{S\arabic{equation}}
\setcounter{figure}{0} 
\setcounter{equation}{0} 
\setcounter{page}{1}
\newpage

\section{Nonhyperbolicity of ($U, V$) flowers}
\label{sec:nonhyperbolicity}

Gromov's $\delta$-hyperbolicity~\cite{bianconi2021higher, gromov1987hyperbolic} provides a convenient quantification of the overall hyperbolicity of a network. It is defined as follows~\cite{bianconi2021higher}: A network is said to be $\delta$-hyperbolic if there exists a finite $\delta$ such that for any triplet of nodes $r$, $s$, and $q$ connected by the
shortest paths $\mathcal{P}_{rs}$, $\mathcal{P}_{sq}$, and $\mathcal{P}_{rq}$ respectively, the union of the $\delta$-neighborhood of any pair of shortest paths includes all nodes that belong to the third shortest path.

Based on this definition, we will show that when $U>1$, the ($U,V$) flowers are not $\delta$-hyperbolic for any finite $\delta$.

\begin{figure}[ht!]
    \centering
    \includegraphics[width=0.8\linewidth]{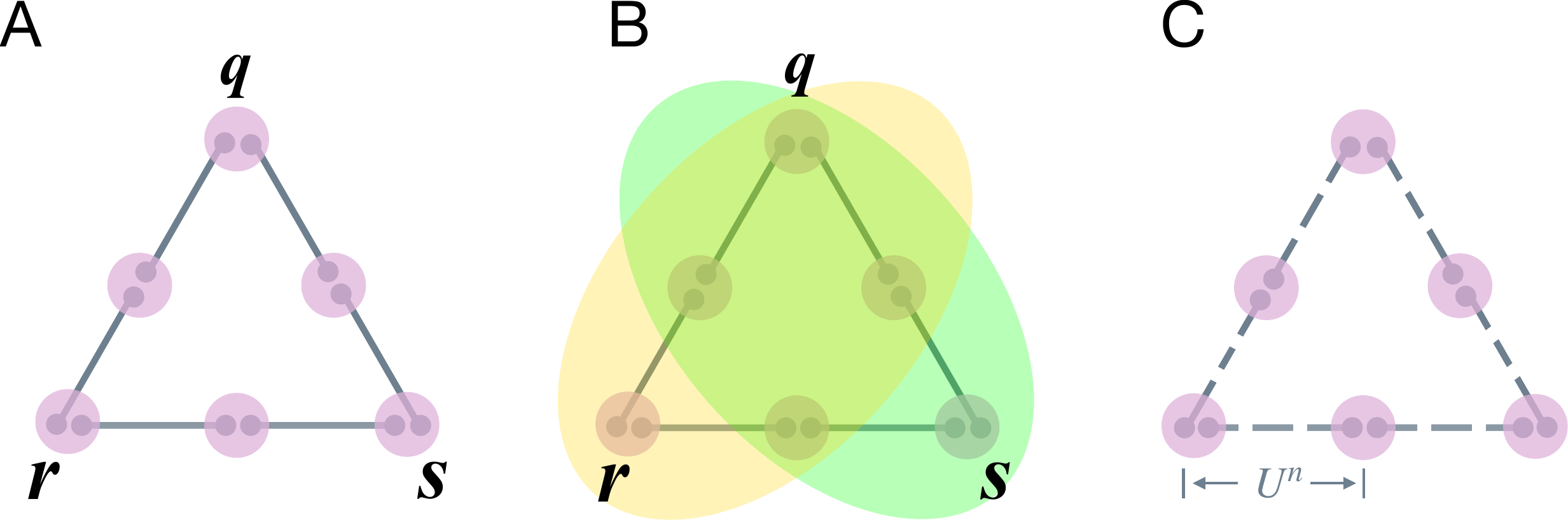}
    \caption{
    \textbf{Proof of nonhyperbolicity of ($U,V$) flower for $U>1$.}
    (\textbf{A}) ($U,V$) flower with $U=2$, $V=4$ of generation $n=1$. Three chosen nodes $q, r, s$ are positioned at a distance of $2$. 
    (\textbf{B}) By choosing $\delta=1$, the $\delta$-neighborhood of the paths  $\mathcal{P}_{rq}$ and $\mathcal{P}_{sq}$ (yellow and green shades) includes all the nodes in the third path, $\mathcal{P}_{rs}$. 
    (\textbf{C}) In the $\left(n+1\right)$-th generation, 
    each link (solid line) in (\textbf{A}) is replaced by an $n$-th generation of the ($U,V$) flower (dashed line) by iteration. This means that the shortest distance between $r$ and $s$ in (\textbf{A}) changes from $2$ to $2 U^n$, which is much greater than $2$ for large $n$ and $U>1$.
    \hfill\hfill}
    \label{hyperbolic}
\end{figure}

Let us consider three nodes  $r, s, q$ in the ($U,V$) flower at generation $n=1$, such that the three nodes are equidistant, each separated by a distance of $(U+V)/3$. As a result, the corresponding three shortest paths $\mathcal{P}_{rs}$, $\mathcal{P}_{rq}$, and $\mathcal{P}_{sq}$ do not overlap with each other {(Fig.~\ref{hyperbolic}A)}. In order to include nodes belonging to the third path, say $\mathcal{P}_{rs}$, within the $\delta$-neighborhood of the union of the other two paths $\mathcal{P}_{rq}$ and $\mathcal{P}_{sq}$, $\delta$ must satisfy
\begin{equation}
    \delta \geq \max_{i \in \mathcal{P}_{rs}} \left\{\min\left\{ \min_{j \in \mathcal{P}_{rq}}d(i, j), \min_{k \in \mathcal{P}_{sq}} d(i, k)\right\}\right\},
    \label{eq:max_min_min}
\end{equation}
where $d(i,j)$ indicates the distance between node $i$ and $j$. 
Since the shortest path between node pairs in $\mathcal{P}_{rs}$ and  $\mathcal{P}_{qr}$ (or $\mathcal{P}_{qs})$ must pass through node $q$ (or $s$), therefore,
\begin{equation}
   \min_{j \in \mathcal{P}_{rq}} d(i, j) = d(i,r), \quad  \min_{k \in \mathcal{P}_{sq}} d(i, k) = d(i,s).
\end{equation}
When $i$ is the midpoint of $r$ and $s$, the quantity on the right-hand side of Eq.~\eqref{eq:max_min_min} is maximized, which is equal to (Fig.~\ref{hyperbolic}B)
\begin{equation}
     \delta \geq \max_{i \in \mathcal{P}_{rs}} \left\{\min\left\{d(i,r), d(i,s) \right\} \right\}= \frac{1}{2} \frac{U+V}{3}.
\end{equation}
Moving to the $\left(n+1\right)$-th generation, since every two neighboring nodes in the first-generation ($U,V$) flower will now be separated by a shortest distance $U^n$ {(Fig.~\ref{hyperbolic}C)}, $\delta$ at the $\left(n+1\right)$-th generation must satisfy
\begin{equation}
\delta \geq \frac{1}{2} \frac{U+V}{3} U^n,
\end{equation}
which is unbounded at $n \to \infty$. Thus, general ($U, V$) flowers with $U>1$ are not hyperbolic. For the special case of $U=1$, the ($U,V$) flowers reduce to the pseudofractal simplicial and cell complexes, which have been shown to be hyperbolic~\cite{bianconi2021higher}.

\section{Critical conditions of classical and concurrence percolation transitions}

The critical threshold $p_\text{th}$ ($c_\text{th}$) 
can be obtained by analyzing the stability of the fixed points $x=x^*$ that satisfy the exact renormalization-group (RG) equation:
\begin{equation}
    \mathcal{R}(x)=x
\end{equation}
where
\begin{eqnarray}
\label{eq_si_R}
\mathcal{R}(x)=\text{para}(\text{seri}(\stackrel{U}{\overbrace{x,x,\dots,x}}),\text{seri}(\stackrel{V}{\overbrace{x,x,\dots,x}})).
\end{eqnarray}
Here, for simplicity, we denote both the probability $p$ and the concurrence $c$ by $x$.
The stability can be determined by the derivative of $\mathcal{R}(x)$ at $x=x^*$. If $|\mathcal{R}^\prime(x^*)|<1$, the fixed point $x^*$ is stable;  if $|\mathcal{R}^\prime(x^*)|>1$, the fixed point $x^*$ is unstable.
We will show that both classical and concurrence ``quantum'' percolation on ($U,V$) flowers have only one critical threshold $p_\text{th}$ (or $c_\text{th}$).

We have, for classical percolation,
\begin{equation}
    \mathcal{R}(p) = 1 - \left(1-p^U\right)\left(1-p^V\right);\label{eq:Rp}
\end{equation}
and for concurrence percolation,
\begin{equation}
    \mathcal{R}(c) = \sqrt{1-\left(2K(c)-1\right)^2},\label{eq:Rc}
\end{equation}
where
\begin{equation}
    K(c) = \max\left(\frac{1+\sqrt{1-c^{2U}}}{2} \frac{1+\sqrt{1-c^{2V}}}{2}, \frac{1}{2}\right).
\end{equation}
When $U=1$, for both classical and concurrence percolation, $\mathcal{R}(x)=x$ only has two trivial fixed points $x^*=0$ and $x^*=1$, and only $x^*=1$ is stable. 
This indicates a critical threshold $x_\text{th}=0$.

\begin{figure}[ht!]
    \centering
    \includegraphics[width=\linewidth]{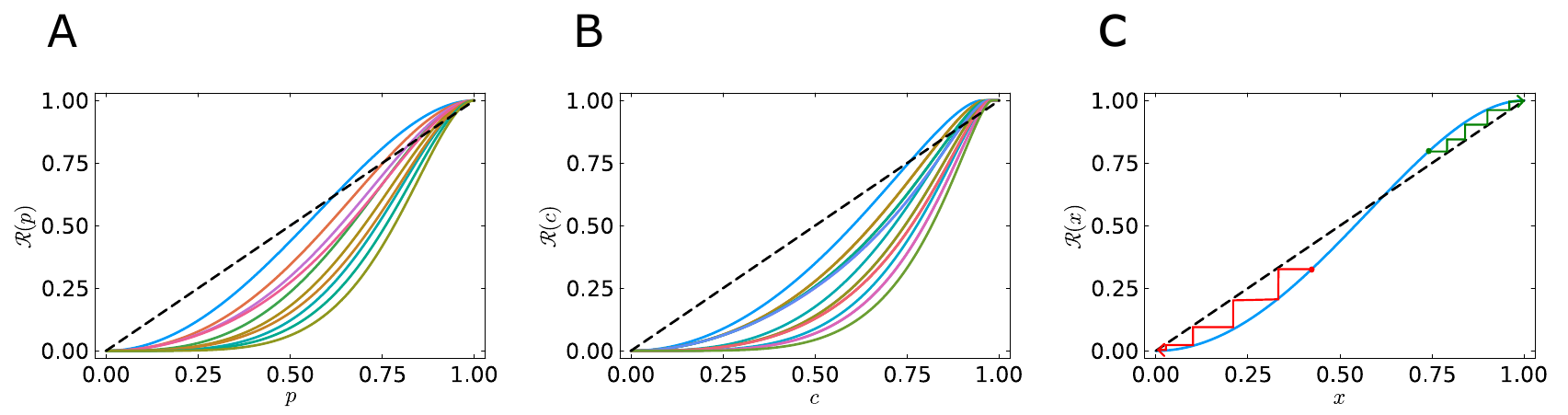}
    \caption{\textbf{Illustration of fixed points of $\mathcal{R}(x)=x$ and their stability.} (\textbf{A}) The map $\mathcal{R}(p)$ defined in Eq.~\eqref{eq:Rp} as a function of $p$. (\textbf{B}) $\mathcal{R}(c)$ defined in Eq.~\eqref{eq:Rc} as a function of $c$. Both maps are shown for $U, V=2, 3, 4, 5$. (\textbf{C}) Cobweb plot illustrating the stability of the fixed points. For any $x<x_\text{th}$ (the red point), $x \in \{c, p\}$, the iterative map will converge to zero, while for any $x>x_\text{th}$, the iterative map will converge to one. The iterative calculations are graphically shown as the red and green lines.\hfill\hfill}
    \label{fig:single_nontrivial_root}
\end{figure}

When $U>1$, for both classical and concurrence percolation, $\mathcal{R}(x)=x$ has three fixed points: $x^*=0$, $x^*=1$, and a nontrivial fixed point $0 < x^* < 1$ (Figs.~\ref{fig:single_nontrivial_root}A~and~\ref{fig:single_nontrivial_root}B).

We observe that $x^*=0$ and $x^*=1$ are two stable fixed points, since $\mathcal{R}^\prime(0)=\mathcal{R}^\prime(1)=0$, and the nontrivial fixed point is always unstable (Fig.~\ref{fig:single_nontrivial_root}C). Thus, the nontrivial fixed point gives the critical threshold $x_\text{th}$: 
iterating Eq.~\eqref{eq_si_R} infinite times yields either $0$ for $x<x_\text{th}$ or $1$ for $x>x_\text{th}$.
Therefore, the percolation threshold for both classical and concurrence percolation is unique.

\section{A noncluster definition of percolating strength}
\begin{figure}[ht!]
    \centering
  \includegraphics[width=\linewidth]{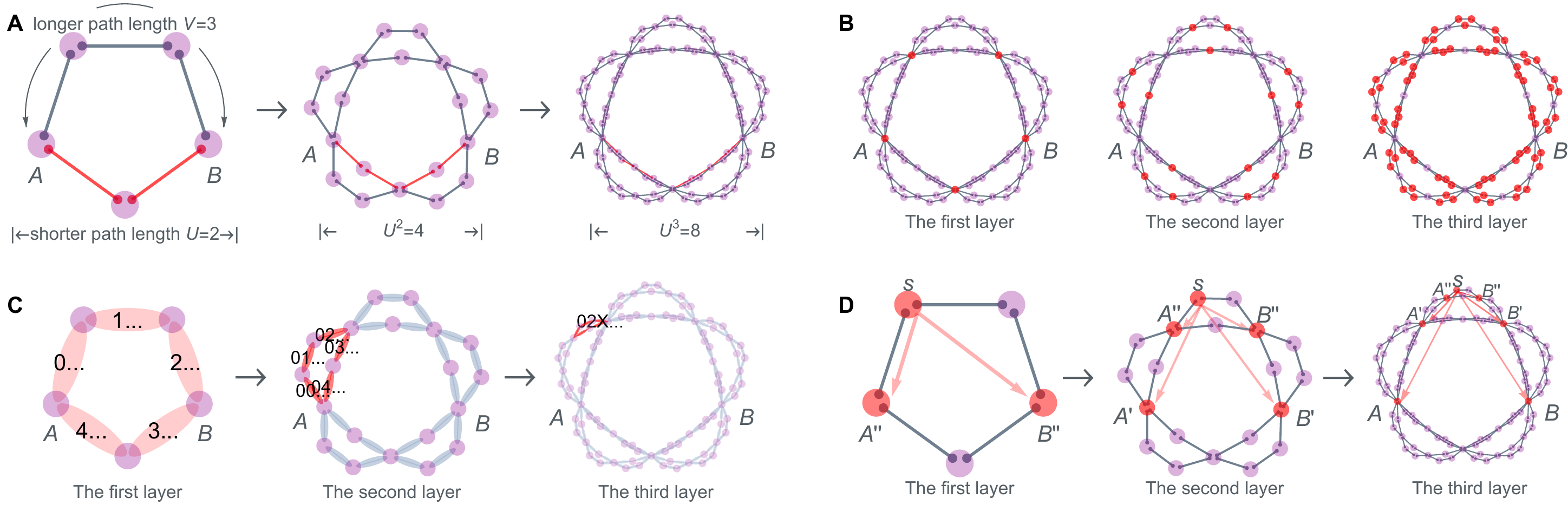}
    \caption{\textbf{Hierarchical scale-free networks and noncluster definition of $P_\infty$.} (\textbf{A})~The construction of hierarchical scale-free quantum networks ($U,V$) flowers where $U>1$ is the shorter length scale while $V\geq U$ is the longer length scale. (\textbf{B})~Nodes on different layers. The deepest (third) layer has the most number of nodes. (\textbf{C})~A node has the same probability of belonging to one of these ($U+V$) regimes, for each layer. (\textbf{D})~We can find a transfer matrix demonstrating the probability that a node connecting to \textit{A}$'$ (or \textit{B}$'$) at layer $2$ in terms of the probability that this node connecting to \textit{A}$''$ (or \textit{B}$''$) at layer $1$, or the probability that it connecting to \textit{A} (or \textit{B}) at layer $3$ in terms of the probability that this node connecting to \textit{A}$'$ (or \textit{B}$'$) at layer $2$.
    \hfill\hfill}
    \label{23flowerPinf}
\end{figure}

First, we categorize the nodes in the ($U,V$) flower into different layers based on the generation in which they were introduced into the network (Figs.~\ref{23flowerPinf}A~and~\ref{23flowerPinf}B). The deepest layer consists of all nodes with the lowest degree of 2, representing the nodes introduced in the last generation. Note that the number of nodes introduced in the ($U,V$) flower increases exponentially with each generation. Consequently, the probability that a node in the deepest layer connects to \textit{A} or \textit{B} governs \emph{the probability of a randomly chosen node in the bulk to reach the boundaries (\textit{A} or \textit{B}) of the network}, which denotes the strength of the percolating cluster, $P_\infty$~\cite{Fractal_PRE}.

However, not all nodes at the deepest layer are equivalent: some are closer to \textit{A} (or \textit{B}) and some are farther. When $n\to \infty$, each node at the deepest layer can be characterized by an infinite series of ($U+V$) digits: For the ($2,3$) flower, for example, each node is characterized by, e.g.,~$434032\dots$, where the $n$-th digit represents which link $(0,1,2,3,4)$ at the $n$-th generation the node traces back to and originates from (Fig.~\ref{23flowerPinf}C). The probability that the node at position, e.g.,~$434032\dots$, connects to \textit{A} (or \textit{B}) is denoted by $P_{434032\dots}$, which is a $2\times 1$ vector: $\text{(prob. to \textit{A}, prob. to \textit{B})}^T$.  The probability of randomly choosing a node at the deepest layer that connects to \textit{A} (or \textit{B}) is, therefore, $P_\infty = P_{0000\dots}+P_{1000\dots}+P_{2000\dots}+\dots+P_{4340\dots}+\dots$ which is also represented by a $2\times 1$ vector.

Now, what is special about classical percolation in the ($U,V$) flower is that $P_{434032\dots}$ can indeed be written as the multiplication of a series of $2\times 2$ matrices and a trivial vector $(1,0)^T$ (the exact form of the trivial vector should not matter): $P_{434032\dots}=\left(P_4 P_3 P_4 P_0 P_3 P_2 \dots \right)(1,0)^T$. Each matrix (e.g.,~$P_0$) is a function of $p$. The four entries of $P_0$ correspond to the conditional probabilities of a node connecting to \textit{A} (\textit{B}) at the $n$-th layer given the probability of the same node connecting to \textit{A}$'$ (\textit{B}$'$) at the $\left(n+1\right)$-th layer (Fig.~\ref{23flowerPinf}D). In other words, $P_0$ plays the role of the transfer matrix in statistical physics, converting from the $\left(n+1\right)$-th to the $n$-th layer, provided that the node traces back to link $0$ at the $\left(n+1\right)$-th generation. By averaging over the five possible digits $(0,1,2,3,4)$, we can rewrite $P_\infty$ as: 
\begin{equation}
\label{eq_p_inf_path}
    \begin{aligned}
        P_\infty &= [\frac{P_0+P_1+P_2+P_3+P_4}{5}][\frac{P_0+P_1+P_2+P_3+P_4}{5}]\dots
        \\
        &= \bar{P} \bar{P} \dots
    \end{aligned}
\end{equation}
This allows us to solve $\beta$ accurately, which is derived from the largest eigenvalue of $\bar{P}=\left({P_0+P_1+P_2+P_3+P_4}\right)/5$. The result is identical to Rozenfeld and ben-Avraham’s calculation~\cite{RHD_PRE2}. 

What complicates the calculation for concurrence percolation is that we cannot write down $C_{434032\dots}$ in terms of a series of transfer matrices. Instead, we write down $C_{434032\dots}$ in terms of a series of transfer functions, given by $C_{434032\dots} = C_4(C_3(C_4(C_0(C_3(C_2)))))\dots$ where each function (e.g.,~$C_0$) has $2\times 1$ input and $2\times 1$ output. The exact form of the function depends on not only the series and parallel rules but also higher-order rules. 

Now, we may nominally define the percolating strength as
\begin{equation}
\label{eq_c_inf_path}
    \begin{aligned}
        C_\infty=\frac{C_0+C_1+C_2+C_3+C_4}{5}(\frac{C_0+C_1+C_2+C_3+C_4}{5}(\dots))
    \end{aligned}.
\end{equation}
Here, $\left[\left({C_0+C_1+C_2+C_3+C_4}\right)/5\right](\cdot)$ denotes the average of the output values of the five different functions, $C_0(\cdot),\dots,C_4(\cdot)$.
This allows us to simulate $C_\infty$ for $n$ layers, arriving at a numerical value of $\beta$ by fitting to the power law.

\section{Star-mesh transform}

The challenge of using Eq.~\eqref{eq_p_inf_path} or Eq.~\eqref{eq_c_inf_path} to calculate the noncluster-defined percolation strength is that the matrices $P_0 \dots$ (or the functions $C_0 \dots$) depend not only on the series and parallel rules but also on higher-order connectivity rules, which cannot be decomposed into series and parallel rules~\cite{conpt_mgh21}. For classical percolation, the higher-order connectivity rules are known, but are highly complicated; for concurrence percolation, these rules are simply unknown. This calls for an approximating approach to treat the higher-order rules as approximate combinations of series and parallel rules only. The star-mesh transform serves for this purpose.

\subsection{Definition}
\begin{figure}[ht!]
	\centering
    \includegraphics[width=\linewidth]{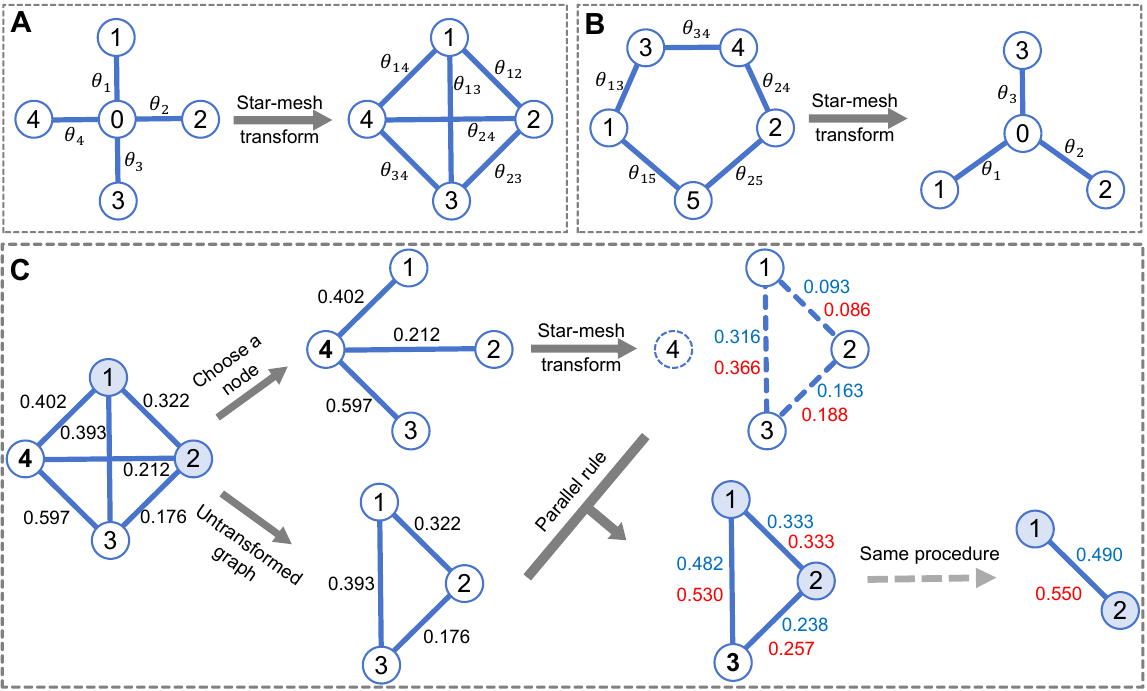}
    \caption{
\textbf{Star-mesh transform.}
(\textbf{A})~Star-mesh transform from an $(s+1)$-node star graph to an $s$-node complete graph. Applying series and parallel rules recursively, the link weights are determined by solving a double-recursive system of $s(s-1)/2$ equations. 
(\textbf{B})~Star-mesh transform from a graph with loop to a star graph. The final weight between same two nodes of the two graphs should be consistent.
(\textbf{C})~Detailed procedure of solving the net weight across, e.g.,~nodes 1 and 2, by reducing the $(s-1)$-node complete graph (blue: classical percolation; red: concurrence percolation).    
\hfill\hfill}
\label{fig_sm}
\end{figure}

A star-mesh transform~\cite{star-mesh_v70} can establish a local equivalence of the connectivity between an $(s+1)$-node star graph and an $s$-node complete graph (Fig.~\ref{fig_sm}A) based only on series and parallel rules~\cite{conpt_mgh21}. We denote the $(s+1)$-node star graph as $\mathcal{G}(s)$, with one root node and $s$ leaf nodes where the weight of the $i$-th link is $\theta_i$. Correspondingly, the star-mesh transform of $\mathcal{G}(s)$, i.e.,~the $s$-node complete graph, is denoted as $\mathcal{G}'(s)$, where the weights of the $s(s-1)/2$ links are assumed as $(\theta_{12},\theta_{13},\dots,\theta_{1s},\dots,\theta_{s-1,s})$.
The equivalence between $\mathcal{G}(s)$ and $\mathcal{G}'(s)$ are given by $s(s-1)/2$ independent equations:
\begin{eqnarray}
\label{star-mesh-eqns}
\text{seri}\left(\theta_1,\theta_2\right) &=&\text{cross}\left(1,2;\mathcal{G}'\left(s\right)\right),\nonumber\\
\text{seri}\left(\theta_1,\theta_3\right) &=&\text{cross}\left(1,3;\mathcal{G}'\left(s\right)\right),\nonumber\\
&\dots&\nonumber\\
\text{seri}\left(\theta_1,\theta_s\right) &=&\text{cross}\left(1,s;\mathcal{G}'\left(s\right)\right),\nonumber\\
&\dots&\nonumber\\
\text{seri}\left(\theta_{s-1},\theta_s\right) &=&\text{cross}\left(s-1,s;\mathcal{G}'\left(s\right)\right).
\end{eqnarray}
The expression $\text{seri}(\theta_i,\theta_j)$ here is the series-sum (based on the series rule) of the $i$-th and $j$-th links in $\mathcal{G}(s)$, and $\text{cross}(i,j;\mathcal{G}'(s))$ is the net weight across the two nodes $i$ and $j$ in $\mathcal{G}'(s)$ and can be calculated by recursively degrading $\mathcal{G}'(s)$ to a link between $i$ and $j$. 

Particularly, first we arbitrarily choose a node (except $i$ or $j$) from $\mathcal{G}'(s)$ as the root of an extracted sub-star-graph $(\text{sub}\mathcal{G}')(s-1)$ which has $s-1$ links connected to the root. Using the star-mesh transform, this sub-star-graph are mapped into an $(s-1)$-complete graph denoted by $(\text{sub}\mathcal{G}')'(s-1)$.  Then we combine it (based on the parallel rule) with the complement subgraph of $\mathcal{G}'(s)$,  denoted by {$\mathcal{G}'(s)\setminus\nobreak(\text{sub}\mathcal{G}')(s-\nobreak1)$},  getting a new $(s-1)$-complete graph. We define the new graph as $\text{Comb}\left(\mathcal{G}_{\alpha},\mathcal{G}_{\beta}\right)$, which has each of its link weight given by $\theta_{ij}=\text{para}(\alpha_{ij},\beta_{ij})$ with $ \alpha_{ij}\in\mathcal{G}_{\alpha}$ and $\beta_{ij}\in\mathcal{G}_{\beta}$. Thus, the connectivity between $i$ and $j$ in $\mathcal{G}'(s)$ can be calculated by 
\begin{equation}
\label{c-recur}
\text{cross}(i,j;\mathcal{G}'(s))=\text{cross}(i,j;\text{Comb}\left((\text{sub}\mathcal{G}')'(s-1),\mathcal{G}'(s)\setminus(\text{sub}\mathcal{G}')(s-1)\right)).
\end{equation}

Next, we choose the nodes of $\mathcal{G}'(s)$ one after the other and operate the same process, until only nodes $i$ and $j$ and a link between them are left, finally producing a function of $\text{cross}(i,j;\mathcal{G}'(s))$ related to $(\theta_{1}, \dots, \theta_{s})$. The entire procedure involves a $(s-1)$-level star-mesh transform, thus is a double recursion (Fig.~\ref{fig_sm}C). Since for concurrence percolation, a closed-form solution of Eqs.~\eqref{star-mesh-eqns} have not been found, we used the Broyden's root-finding algorithm to numerically find the $s(s-1)/2$ weights $\theta_{ij}$ that satisfy Eqs.~\eqref{star-mesh-eqns}. 

The star-mesh transform is similar to the real-space renormalization group (RG) for percolation theory, but is more general and suits for any networks, not just lattices. By consecutively applying it on transforming a star graph to a complete graph where a node is reduced for each time, which allows the degradation of network structure, the connectivity between two nodes \textit{A} and \textit{B} can be well approximated by the final weight $\theta$ of the link between them. 

Alternatively, it can also be used to convert a cycle to a star graph. Based on the connectivity between same two nodes $i$ and $j$, the link weights of this transformed star graph $\left(\theta_1,\theta_2,\dots\right)$ is solved by the equations with the weights of the original graph. For example, in Fig.~\ref{fig_sm}B, the group of equations is:
\begin{subequations}
\begin{eqnarray}
\label{mesh-star-eqns}
    \text{seri}\left(\theta_1,\theta_2\right)
    &=&\text{para}\left(\text{seri}\left(\theta_{15},\theta_{25}\right),\text{seri}\left(\theta_{13},\theta_{34},\theta_{24}\right)\right),\\
    \text{seri}\left(\theta_1,\theta_3\right)
    &=&\text{para}\left(\theta_{13},\text{seri}\left(\theta_{15},\theta_{25},\theta_{24},\theta_{34}\right)\right),\\
    \text{seri}\left(\theta_2,\theta_3\right)
    &=& \text{para}\left(\text{seri}\left(\theta_{24},\theta_{34}\right),\text{seri}\left(\theta_{13},\theta_{15},\theta_{25}\right)\right).
\end{eqnarray}
\end{subequations}
We will use the star-mesh transform to calculate and compare the noncluster-defined $P_\infty$ and $C_\infty$ in the following sections.

\subsection{ Validity of the star-mesh transform}

Since for classical percolation, the exact cluster-defined $P_\infty$ is known, this allows us to  calculate the noncluster-defined, classical $P_\infty$ using the star-mesh transform for different network topologies, testing the validity of the approach by comparing to exact results.

Firstly, we  compare the exact result of classical percolation [$\beta=0.165\dots$ for ($2,2$) flowers] with the star-mesh-transform result of classical percolation [$\beta\approx 0.168829(3)$, which is approximate]. Considering that the star-mesh transform is a type of RG, the result is very plausible for investigating ordinary second-order phase transitions.

Next, to further assess the accuracy and generality of the star-mesh transform, here we extend to hyperbolic networks, focusing on the ($1, V$) flowers which, as noted in Section~\ref{sec:nonhyperbolicity}, are hyperbolic and differ from general ($U,V$) flowers with $U>1$. 
It is worth noting that ($1, V$) flowers can be equivalently treated as pseudofractal simplicial and cell complexes formed by 
$\left(V+1\right)$-sided polygons~\cite{sun2020renormalization}.
This enables a cluster-based, analytical calculation of the peculiar critical behavior of 
$P_\infty$
near the trivial critical threshold, $p_\text{th}=0$. It can be shown that the critical behavior follows a special scaling form~\cite{sun2020renormalization}:
\begin{equation}
\label{eq_1v_p_inf}
    P_\infty \propto \left(p-p_\text{th}\right) \exp\left[-\alpha/\left(p-p_\text{th}\right)^{V-1}\right].
\end{equation}

\begin{figure}[ht!]
   \centering
    \includegraphics[width=\linewidth]{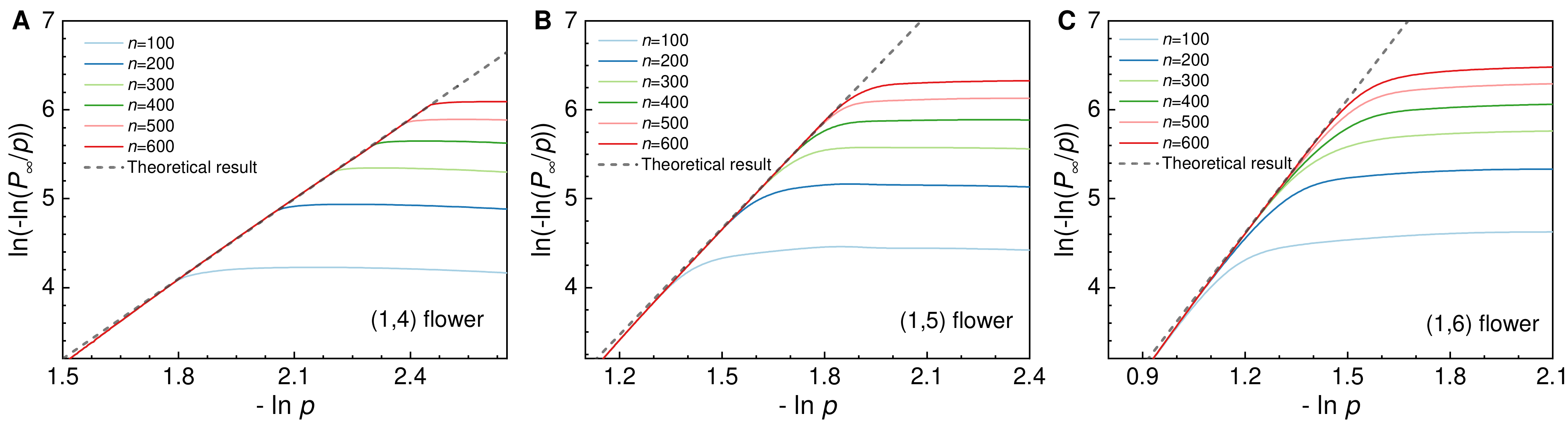}
    \caption{\textbf{Critical scaling of $P_{\infty}$ as a function of $p$ for ($1,V$) flowers.} This figure shows the critical behavior following a special scaling of $P_{\infty} \propto p\exp{(-\alpha / p^{V-1})}$ yielding $\ln(-\ln (P_\infty / p))=\ln \alpha-(V-1)\ln p$ (see Ref.~\cite{sun2020renormalization}) for different (\textbf{A})~$V=4$, (\textbf{B})~$V=5$ and (\textbf{C})~$V=6$. In infinite network limit (dashed line), the scaling is predicted by analyzing the critical behavior of system size $N \sim (V+1)^n$ with $n = 100, 200, \dots, 600$.\hfill\hfill}
    \label{fig_1Vflower}
\end{figure}

In Fig.~\ref{fig_1Vflower}, instead of using the cluster-based definition of $P_\infty$, we calculate $P_\infty$ for ($1, V$) flowers of different $V$ and 
system size $N$ using the star-mesh transform  (colored lines) and compare the numerical results with the exact scaling form, Eq.~\eqref{eq_1v_p_inf} (dashed line). Notably, the star-mesh transform results asymptotically agree with the scaling form.
In addition, the finite-size effect of the saturation of $P_\infty$ for $p \to p_\text{th}$ (or $-\ln p \to +\infty$, as in Fig.~\ref
{fig_1Vflower}) matches our theoretical understanding~\cite{sun2020renormalization},
suggesting high accuracy of our star-mesh transform method.

We further apply the star-mesh transform to the Farey graphs, another type of hyperbolic networks~\cite{boettcher2012ordinary}. At the critical threshold, $p_\text{th}=1/2$, the percolating strength has been shown to follow~\cite{boettcher2012ordinary}
\begin{equation}
   N_g \equiv N P_\infty = {N^{d_f(p)/d}},
\end{equation}
where the fractal dimension of the percolating cluster  now depends on $p$:
\begin{equation}
\label{eq_d_f_p}
    d_f (p)/d \simeq 1-\frac{8(p-p_\text{th})^2}{\ln 2}
\end{equation}
near $p \to p_\text{th}^{-}$.

\begin{figure}[ht!]
   \centering
    \includegraphics[width=0.5\linewidth]{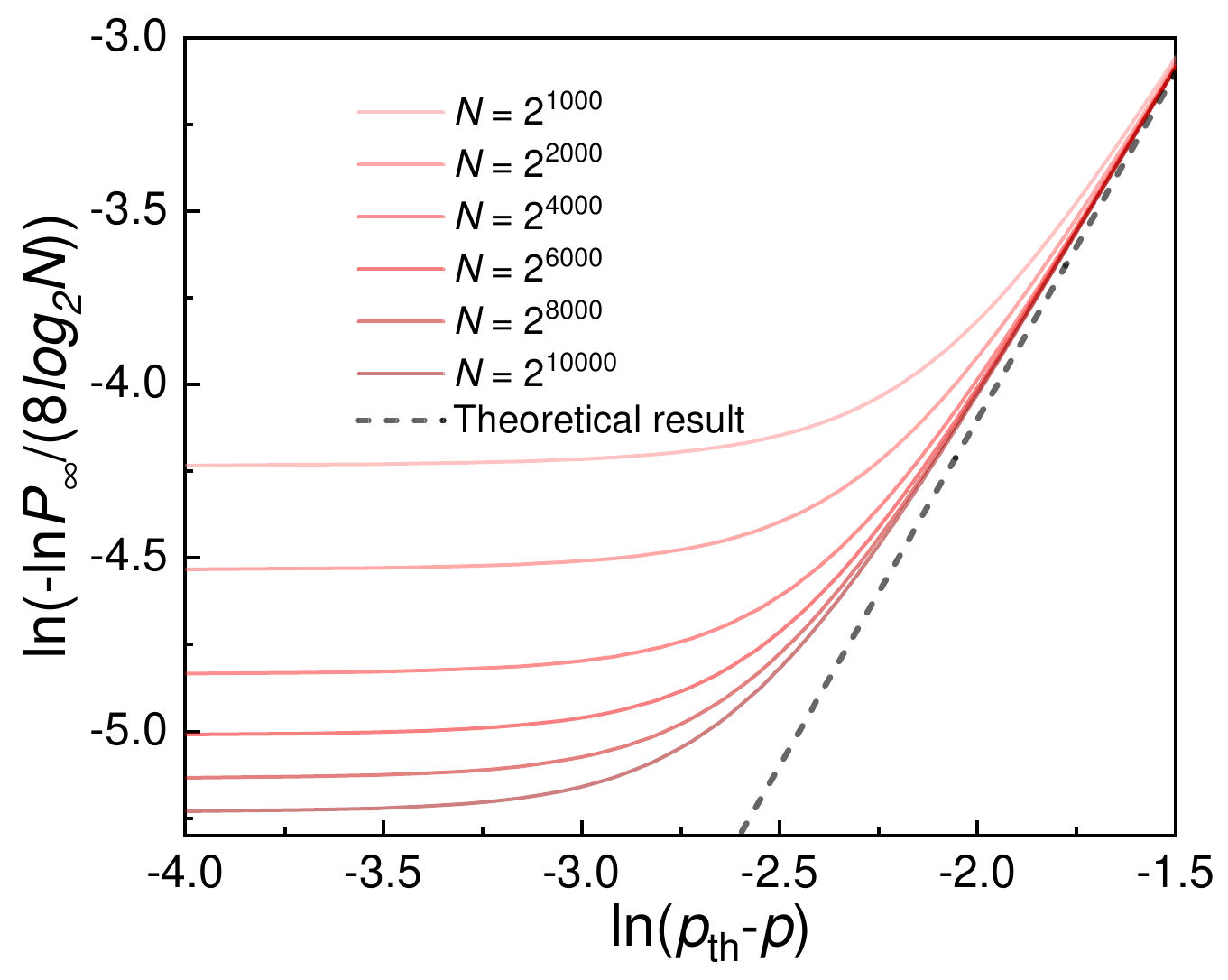}
    \caption{
    \textbf{Finite size scaling of $P_{\infty}$ as a function of $p_\text{th}-p$.} Red lines from lightest to darkest represent network size from $N = 2^{1000}$ to $N= 2^{10000}$. 
    Dashed line denotes the theoretical prediction $P_\infty = N^{d_f(p)/d-1}$, where $d_f(p)/d-1 \simeq -8(p-p_\text{th})^2/\ln{2}$. \hfill\hfill}
    \label{fig_NC}
\end{figure}

In Fig.~\ref{fig_NC}, we calculate $P_\infty$ for different 
$N$, again using the noncluster definition based on the star-mesh transform. Here, again, we observe not only the asymptotic agreement of the scaling form [Eq.~\eqref{eq_d_f_p}] but also the correct finite-size effect.

Taken together, the star-mesh transform is not only useful for calculating the noncluster definition of percolation strength for ordinary second-order phase transitions but also versatile enough for investigating more peculiar critical behaviors. This will be crucial for  studying concurrence percolation (where clusters are not well defined) in hyperbolic networks in the future.

\section{Hyperscaling relation test}
\begin{figure}[ht!]
   \centering
    \includegraphics[width=\linewidth]{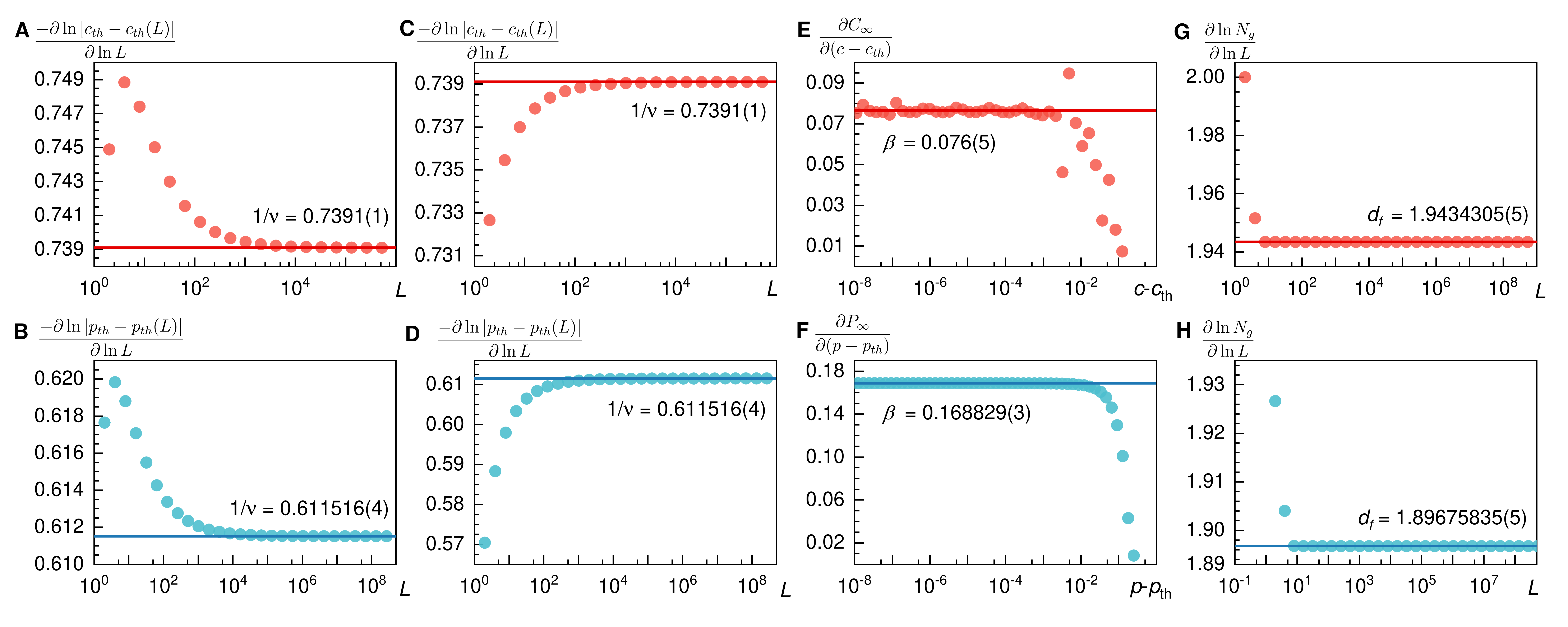}
    \caption{\textbf{Numerical calculations of critical exponents.} (\textbf{A--D})~Numerical determination of $\nu$. (\textbf{A})~quantum $c_{\text{th}}(L)< c_{\text{th}}$ (red) when $n\to \infty$, compared to (\textbf{B})~classical $p_{\text{th}}(L)< p_{\text{th}}$ (blue). Similarly, 
    (\textbf{C})~quantum $c_{\text{th}}(L)> c_{\text{th}}$ (red),
    compared to (\textbf{D})~classical $p_{\text{th}}(L)> p_{\text{th}}$ (blue).
    (\textbf{E--F}) Numerical determination of $\beta$ for (\textbf{E})~concurrence percolation (red) and 
    (\textbf{F})~classical percolation (blue).
    (\textbf{G--H}) Using the data from Figs.~\ref{fig_percolations}(I)~and~\ref{fig_percolations}(J), we extract the local slope $\partial{\ln{N_g}} / \partial{\ln{L}}$ in classical percolation and concurrence percolation, respectively, to determine the fractal dimension $d_f$.\hfill\hfill}
    \label{fig_exponents}
\end{figure}

The alternative, noncluster definition of $P_\infty$ and $C_\infty$ leads to a numerical method of calculating the critical exponents. Here we focus on ($2,2$) flowers. Let $L\to \infty$, it can be easily derived from the traditional relations. More intuitively, for classical percolation $ 1 / \nu \simeq -\partial{(|p_{\text{th}}(L)-p_{\text{th}}|)} / \partial{L}$ yielding $1 / \nu \approx 0.61152$ while for concurrence percolation $ 1 / \nu \simeq -\partial{(|c_{\text{th}}(L)-c_{\text{th}}|)} / \partial{L}$ yielding $1 / \nu \approx 0.73911$ (Figs.~\ref{fig_exponents}A--D), which are in very good agreement with the analytical result. And $\beta \simeq -\nu \partial{\ln{P_\infty}} / \partial{\ln{L}}$ or $\beta \simeq -\nu \partial{\ln{C_\infty}} / \partial{\ln{L}}$ (Figs.~\ref{fig_exponents}E--F), while $d_f \simeq \partial{\ln{N_g}} / \partial{\ln{L}}$ (Figs.~\ref{fig_exponents}G--H).

\section{Asymptotic analysis of critical phenomena on ($U,V$) flowers}

In the following, we calculate the critical thresholds $p_{\text{th}}$ ($c_{\text{th}}$), percolating strengths $P_\infty$ ($C_\infty$), and critical exponents ($\nu$, $\beta$, $d_f$) in details, with a focus on the asymptotic limit $V\to \infty$.

\subsection{\, Classical percolation}
\subsubsection{Critical threshold $p_{\text{th}}$}
We solve the fixed point equation of the RG function:
\begin{equation}
    \begin{aligned}
        \mathcal{R}(p) &= \text{para}\left(\text{seri}\stackrel{U}{\left(\overbrace{p,p,\dots,p}\right)},\text{seri}\stackrel{V}{\left(\overbrace{p,p,\dots,p}\right)}\right) \\
        &= 1-(1-p^U)(1-p^V)\\
        &= p^U + p^V - p^{U+V}\\
        &=p,
    \end{aligned}
\end{equation}
which produces
\begin{equation}
    p_{\text{th}} \simeq 1-AV^{-1}+O(V^{-2}) \quad \text{as }V \to \infty,\quad A=\ln{\frac{U}{U-1}} \simeq \frac{1}{U}.
\end{equation}

\subsubsection{Critical exponent $\nu$}
In order to find the critical exponent $\nu$, we evaluate
\begin{equation}
\label{Lamuda}
    \frac{\partial \mathcal{R}(p)}{\partial p}\big|_{p=p_{\text{th}}} = \frac{\partial (p^U + p^V - p^{U+V})}{\partial p}\big|_{p=p_{\text{th}}}.
\end{equation}
Additionally, we have
\begin{equation}
    \lim\limits_{{V \to \infty}} p_{\text{th}}^U \simeq \lim\limits_{{V \to \infty}} (1-AV^{-1})^U \simeq 1-UAV^{-1}
\end{equation}
and
\begin{equation}
    \lim\limits_{{V \to \infty}} p_{\text{th}}^V = \lim\limits_{{V \to \infty}} (1-AV^{-1})^V = e^{-A} = \frac{U-1}{U},
\end{equation}
which correspond to the sponge-crossing connectivity \emph{at criticality} along the shorter path and the longer path, respectively.

Substituting $p_{\text{th}}^U$ and $p_{\text{th}}^V$ into Eq.~\eqref{Lamuda}, we have
\begin{equation}
    \begin{aligned}
        \frac{\partial \mathcal{R}(p)}{\partial p}\big|_{p_{\text{th}}} &= p^{-1}(Up^U + Vp^V -(U+V)p^{U+V})\big|_{p=p_{\text{th}}} \\
        &= p^{-1}(Up^U(1-p^V) + Vp^V(1-p^U))\big|_{p=p_{\text{th}}} \\
        &\simeq (1-AV^{-1})^{-1}(U(1-UAV^{-1}) + Ve^{-A}(1-(1-UAV^{-1})) - U(1-UAV^{-1})e^{-A}) \\
        &\simeq (1+AV^{-1})(U(1-UAV^{-1})(1-e^{-A}) + UAe^{-A}) \\
        &= \left(1-e^{-A}+Ae^{-A}\right)U + \left((1-e^{-A}+Ae^{-A})UA - (1-e^{-A})U^2A\right)V^{-1} \\
        &\quad\; - \left((1-e^{-A})U^2A^2\right)V^{-2}.
    \end{aligned}
\end{equation}
So the critical exponent $\nu$ is given by
\begin{equation}
    \begin{aligned}
    \nu &= \frac{\ln U}{\ln \left( \dfrac{\partial \mathcal{R}(p)}{ \partial p} \big|_{p=p_{\text{th}}}\right)}\\
    &= \frac{\ln U}{\ln U + \ln (1-e^{-A}+Ae^{-A})} + O(V^{-1}) \\
    &= \frac{\ln U}{\ln (1+(U-1)\ln{\frac{U}{U-1}})} + O(V^{-1}).
    \end{aligned}
\end{equation}

\subsubsection{Percolating strength $P_\infty$}
The noncluster-defined $P_\infty$ is solved by:
\begin{equation}
\label{p_inf}
    \begin{cases}
        \left\{
        \begin{aligned}
        \text{seri}(x', t') &= \text{seri}(t, \text{para}(\text{seri}(x,\stackrel{a}{\overbrace{p,p,\dots,p}}),\text{seri}(y,\stackrel{U+V-1-a}{\overbrace{p,p,\dots,p}}))) \\
        \text{seri}(y', t') &= \text{seri}(t, \text{para}(\text{seri}(x,\stackrel{U-1-a}{\overbrace{p,p,\dots,p}}),\text{seri}(y,\stackrel{V+a}{\overbrace{p,p,\dots,p}}))) \\
        \text{seri}(x', y') &= \text{para}(\text{seri}(\stackrel{U}{\overbrace{p,p,\dots,p}}),\text{seri}(\stackrel{V}{\overbrace{p,p,\dots,p}})) 
        \end{aligned}
        \right.
        & \text{, where } a = 0, 1, 2, \dots, U-1;\\
\\
        \left\{
        \begin{aligned}
        \text{seri}(x', t') &= \text{seri}(t, \text{para}(\text{seri}(x,\stackrel{b}{\overbrace{p,p,\dots,p}}),\text{seri}(y,\stackrel{U+V-1-b}{\overbrace{p,p,\dots,p}}))) \\
        \text{seri}(y', t') &= \text{seri}(t, \text{para}(\text{seri}(x,\stackrel{V-1-b}{\overbrace{p,p,\dots,p}}),\text{seri}(y,\stackrel{U+b}{\overbrace{p,p,\dots,p}}))) \\
        \text{seri}(x', y') &= \text{para}(\text{seri}(\stackrel{U}{\overbrace{p,p,\dots,p}}),\text{seri}(\stackrel{V}{\overbrace{p,p,\dots,p}})) 
        \end{aligned}
        \right.
        & \text{, where } b = 0, 1, 2, \dots, V-1.\\
    \end{cases}
\end{equation}
For the classical case, we know that $\text{seri}(p_1,p_2) = p_1 p_2$ and $\text{para}(p_1,p_2) = 1-(1-p_1)(1-p_2) = p_1+p_2-p_1 p_2$. Thus, Eq.~\eqref{p_inf} reduces to
\begin{equation}
\label{u+v equations}
    \begin{cases}
        \left\{
        \begin{aligned}
        x't' &= t(x p^a + y p^{U+V-1-a} - xy p^{U+V-1}) \\
        y't' &= t(x p^{U-1-a} + y p^{V+a} - xy p^{U+V-1}) \\
        x'y' &= p^U + p^V -p^{U+V}
        \end{aligned}
        \right.
        & \text{, where } a = 0, 1, 2, \dots, U-1;\\
\\
        \left\{
        \begin{aligned}
        x't' &= t(x p^b + y p^{U+V-1-b} - xy p^{U+V-1}) \\
        y't' &= t(x p^{V-1-b} + y p^{U+b} - xy p^{U+V-1}) \\
        x'y' &= p^U + p^V -p^{U+V}
        \end{aligned}
        \right.
        & \text{, where } b = 0, 1, 2, \dots, V-1.\\
    \end{cases}
\end{equation}
When $n \to \infty$, we assume $x'=x=y'=y$, thus $xy=\text{para}(p^U,p^V)\stackrel{p\to p_{\text{th}}}=p$ gives $x=y=\sqrt{p}$.
By Eq.~\eqref{u+v equations}, the average value satisfies
\begin{equation}
    \begin{aligned}
        \dfrac{t'}{t} &= \dfrac{\sum_{a=0}^{U-1} \sqrt{(p^{1/2} p^a + p^{U-1/2} p^V p^{-a} - p^U p^V)(p^{U-1/2} p^{-a} + p^{1/2} p^V p^a - p^U p^V)}}{(U+V) \sqrt{p^U+p^V-p^{U+V}}} \\
        &+ \dfrac{\sum_{b=0}^{V-1}\sqrt{(p^{1/2} p^b + p^{U-1/2} p^V p^{-b} - p^U p^V)(p^{V-1/2} p^{-b} + p^{1/2} p^U p^b - p^U p^V)}}{(U+V) \sqrt{p^U+p^V-p^{U+V}}}.
    \end{aligned}
\end{equation}
We denote 
\begin{equation}
    \begin{aligned}
        f_1 & = \sum_{a=0}^{U-1}\sqrt{(p^{1/2} p^a + p^{U-1/2} p^V p^{-a} - p^U p^V)(p^{U-1/2} p^{-a} + p^{1/2} p^V p^a - p^U p^V)} \\
        &\stackrel{V \to \infty}{\simeq} \sum_{a=0}^{U-1}\sqrt{ (1+e^{-A}-e^{-A}) (1+e^{-A}-e^{-A})} \\
        &\stackrel{V \to \infty}{\simeq} U+O(1),
    \end{aligned}
\end{equation}
and
\begin{equation}
    \begin{aligned}
        f_2 &= \sum_{b=0}^{V-1}\sqrt{(p^{1/2} p^b + p^{U-1/2} p^V p^{-b} - p^U p^V)(p^{V-1/2} p^{-b} + p^{1/2} p^U p^b - p^U p^V)} \\
        &\stackrel{V \to \infty}{\simeq} \int_{0}^{V} \sqrt{(p^b + p^{U+V-1-b} - p^{U+V-1}) (p^{V-1-b} + p^{U+b} - p^{U+V-1})} \, db \\
        &= \int_{p^0}^{p^V} \frac{\sqrt{(p^b + p^{U+V-1-b} - p^{U+V-1}) (p^{V-1-b} + p^{U+b} - p^{U+V-1})}}{p^b \ln{b}} \, d(p^b) \\
        &\stackrel{V \to \infty}{\simeq} \int_{1}^{e^{-A}} -\frac{V}{A} x \sqrt{(x + p^{U+V-1} x^{-1} - p^{U+V-1}) (p^{V-1} x^{-1} + p^{U} x - p^{U+V-1})} \, dx \\
        &\stackrel{V \to \infty}{\simeq} \int_{1}^{e^{-A}} -\frac{V}{A}(1 + e^{-A} x^{-2} - e^{-A} x^{-1}) \, dx \\
        &= -\frac{V}{A} (2e^{-A}+Ae^{-A}-2).
    \end{aligned}
\end{equation}
Therefore,
\begin{equation}
    \dfrac{P'_\infty}{P_\infty} = \dfrac{t'}{t} = \dfrac{f_1 + f_2}{(U+V) \sqrt{p^U+p^V-p^{U+V}}}.
\end{equation}

\subsubsection{Critical exponents $d_f$ and $\beta$}
Let $\dfrac{P'_\infty}{P_\infty} = (U+V)^{-\theta}$, where the exponent $\theta$ traditionally characterizes the size of the giant component at the critical threshold~\cite{RHD_PRE2}. When $V \to \infty$, we have
\begin{equation}
    \begin{aligned}
        V^{-\theta} &\simeq -\dfrac{V}{A} \dfrac{2e^{-A}+Ae^{-A}-2}{(U+V)(1+O(V^{-1}))} \\
        &\simeq -\dfrac{\ln{\frac{U}{U-1}}\frac{U-1}{U}-\frac{2}{U}}{\ln{\frac{U}{U-1}}} + O(V^{-1}) \\
        &= - \frac{U-1}{U} + \frac{2}{U} \left(\ln{\frac{U}{U-1}}\right)^{-1},
    \end{aligned}
\end{equation}
producing 
\begin{equation}
    \theta \simeq \dfrac{\ln{\left(\frac{U}{(1-U) + 2(\ln{\frac{U}{U-1}})^{-1}}\right)}}{\ln{V}}.
\end{equation}
The exponent $\theta$ is directly related to $d_f$ by $d_f \equiv d(1-\theta)$, therefore, 
\begin{equation}
    \begin{aligned}
        d_f &\simeq \frac{\ln{V}}{\ln{U}} \left(1-\dfrac{\ln{\left(\frac{U}{(1-U) + 2(\ln{\frac{U}{U-1}})^{-1}}\right)}}{\ln{V}}\right) \\
        &\simeq \frac{\ln{V}}{\ln{U}} - \dfrac{\ln{\left(\frac{U}{(1-U) + 2(\ln{\frac{U}{U-1}})^{-1}}\right)}}{\ln{U}} + O(V^{-1}).
    \end{aligned}
\end{equation}
Assuming the hyperscaling relation $\beta = \nu (d-d_f)$, we derive
\begin{equation}
    \begin{aligned}
        \beta &\simeq \left(\frac{\ln U}{\ln (1+(U-1)\ln{\frac{U}{U-1}})} + O(V^{-1})\right) \left(\frac{\ln{\left(\frac{U}{(1-U) + 2(\ln{\frac{U}{U-1}})^{-1}}\right)}}{\ln{U}} + O(V^{-1})\right) \\
        &\simeq \frac{\ln{\left(\frac{U}{(1-U) + 2(\ln{\frac{U}{U-1}})^{-1}}\right)}}{\ln (1+(U-1)\ln{\frac{U}{U-1}})} + O(V^{-1}).
    \end{aligned}
\end{equation}

\subsection{\, Concurrence percolation}
\subsubsection{Critical threshold $c_{\text{th}}$}
Let $c = 1-\dfrac{m(U,V)V^{-1}}{2}$, thus $c^2 = \left(1-\dfrac{m(U,V)V^{-1}}{2}\right)^2 = 1-m V^{-1}+O(V^{-2})$. We know $\lim\limits_{{V \to \infty}} c^{2V} = \lim\limits_{{V \to \infty}} \left(1-mV^{-1}\right)^{V} = e^{-m}$ and $c^{2U} \simeq (1-mV^{-1})^{U} \simeq 1-mUV^{-1}$. Thus, the quantum threshold $c_{\text{th}}$ is solved by 
\begin{equation}
\label{c_th solution}
    \begin{aligned}
        \frac{1 + \sqrt{1-c^2}}{2} &= \frac{1 + \sqrt{1-c^{2U}}}{2} \cdot \frac{1 + \sqrt{1-c^{2V}}}{2} \Rightarrow \\
        \frac{1 + \sqrt{1-(1-\frac{m}{V})}}{2} &= \frac{1 + \sqrt{1-(1-\frac{m}{V})^U}}{2} \cdot \frac{1 + \sqrt{1-(1-\frac{m}{V})^V}}{2} \Rightarrow \\
        \frac{1}{2}(1+\sqrt{\frac{m}{V}}) &= \frac{1}{4} (1+\sqrt{\frac{mU}{V}})(1+\sqrt{1-e^{-m}}),
    \end{aligned}
\end{equation}
which becomes
\begin{equation}
\frac{1 + \sqrt{\frac{mU}{V}}}{2} \dots \frac{1 + \sqrt{1-e^{-m}}}{2}  = \frac{1 + \sqrt{\frac{m}{V}}}{2} \quad \text{for $V \gg 1$}.
\end{equation}
Noting that $1 + \sqrt{1-e^{-m}} \approx 2 - \frac{1}{2}e^{-m}$, multiplying by 2, taking the logarithm and rearranging yields
\begin{equation}
\ln \left(1 + \sqrt{\frac{m}{V}}\right)
= \ln \left(1 + \sqrt{\frac{mU}{V}}\right) + \ln \left(1 - \frac{1}{4}e^{-m}\right).
\end{equation}
Using $\ln (1 \pm x) \approx \pm x$ for $x \to 0$, we find
\begin{equation}
\sqrt{\frac{m}{V}}
= \sqrt{\frac{mU}{V}} - \frac{1}{4}e^{-m} \quad \text{for $V \gg 1$}
\end{equation}
or 
\begin{equation}
\sqrt{m}
= \sqrt{mU} - \frac{1}{4}e^{\frac{1}{2}\ln V -m} \quad \text{for $V \gg 1$}, \label{Eq:msolXinqi}
\end{equation}
which, divided by $\sqrt{m}$, gives rise to
\begin{equation}
1
= \sqrt{U} - \frac{1}{4}\sqrt{\frac{V}{m}} e^{-m},
\end{equation}
that is,
\begin{equation}
e^{\frac{1}{2}\ln V - m - \frac{1}{2} \ln m} = 4 \left(\sqrt{U} - 1\right).    
\end{equation}
So taking the (natural) logarithm and rearranging, we have
\begin{equation} \label{Eq:solvemV}
m + \frac{1}{2} \ln m - \frac{1}{2}\ln V + \ln \left(4 \left(\sqrt{U} - 1\right)\right) = 0 \quad \text{for $V \gg 1$}.
\end{equation}
Hence, we can find $m$ numerically for fixed $U$ as a function of $V$ by solving this equation or Eq.~\eqref{Eq:msolXinqi}. Note that $V$ only appears via $\ln V$. Hence, there is no numerical problems with solving this equation even for $V \to \infty$. 
Additionally, from Eq.~\eqref{Eq:solvemV} we have
\begin{equation}
    \lim\limits_{{V \to \infty}} c_{\text{th}}^U \simeq 1-\frac{1}{2}m U V^{-1} \simeq 1-\frac{1}{4} U V^{-1} \ln V
\end{equation}
and
\begin{equation}
    \lim\limits_{{V \to \infty}} c_{\text{th}}^V \simeq e^{-m/2} \simeq V^{-\frac{1}{4}},
\end{equation}
which correspond to the sponge-crossing connectivity \emph{at criticality} along the shorter path and the longer path, respectively.

\subsubsection{Critical exponent $\nu$}
We denote $\Lambda = \dfrac{\partial c'}{\partial c}$. When $V \to \infty, \, c'=c$, we have $\Lambda = \dfrac{\partial c'}{\partial c'^2} \dfrac{\partial c'^2}{\partial c^2} \dfrac{\partial c^2}{\partial c} = \dfrac{1}{2c'} \dfrac{\partial c'^2}{\partial c^2} 2c = \dfrac{\partial c'^2}{\partial c^2}$, where we write $\dfrac{\partial C'}{\partial C} = \dfrac{\partial c'^2}{\partial c^2}$. Using $\dfrac{1 + \sqrt{1-c'^2}}{2} = \dfrac{1 + \sqrt{1-c^{2U}}}{2} \cdot \dfrac{1 + \sqrt{1-c^{2V}}}{2}$, we have 
\begin{equation}
    \begin{aligned}
        \dfrac{\partial{\left(\dfrac{1 + \sqrt{1-c'^2}}{2}\right)}}{\partial{c^2}} &= \dfrac{\partial {\left(\dfrac{1 + \sqrt{1-c^{2U}}}{2} \dfrac{1 + \sqrt{1-c^{2V}}}{2} \right)}}{\partial{c^2}}\Leftrightarrow \\
        \dfrac{\partial{\left(\dfrac{1 + \sqrt{1-C'}}{2}\right)}}{\partial{C}} &= \dfrac{\partial {\left(\dfrac{1 + \sqrt{1-C^{U}}}{2} \dfrac{1 + \sqrt{1-C^{V}}}{2} \right)}}{\partial{C}}\Leftrightarrow \\
        -\dfrac{1}{\sqrt{1-C'}} \dfrac{\partial C'}{\partial C} &= \dfrac{-UC^{U-1} \left(1+\sqrt{1-C^V}\right)}{2 \sqrt{1-C^U}} + \dfrac{-VC^{V-1} \left(1+\sqrt{1-C^U}\right)}{2 \sqrt{1-C^V}} \Leftrightarrow\\
        \dfrac{\partial C'}{\partial C} &= \dfrac{C^{-1} \sqrt{1-C'}}{2} \left(\dfrac{U C^{U} \left(1+\sqrt{1-C^V}\right)}{\sqrt{1-C^U}} + \dfrac{V C^{V} \left(1+\sqrt{1-C^U}\right)}{\sqrt{1-C^V}} \right).
    \end{aligned}
\end{equation}
When $V \to \infty$, $\lim\limits_{{V \to \infty}} C^V = \lim\limits_{{V \to \infty}} (1-mV^{-1})^V = e^{-m}$, $C^U \simeq (1-mV^{-1})^U \simeq 1-mUV^{-1}$, so
\begin{equation}
     \Lambda = \frac{1}{2} \left(1+\frac{m}{V}\right) \sqrt{\frac{m}{V}} \left(U\left(1-\frac{mU}{V}\right) \frac{1+\sqrt{1-e^{-m}}}{\sqrt{1-\left(1-\frac{mU}{V}\right)}} + Ve^{-m} \frac{1+\sqrt{1-\left(1-\frac{mU}{V}\right)}}{\sqrt{1-e^{-m}}}\right).
     \label{approLamuda1}
\end{equation}
For large $V$, approximations are required. Using $\sqrt{1-x} \approx 1 - \frac{1}{2}x$ for $x \ll 1$, we have $\sqrt{1-e^{-m}} \approx 1 - \frac{1}{2}e^{-m}$, so
\begin{equation}
\begin{aligned}
        \Lambda &= \frac{1}{2} \left(1+\frac{m}{V}\right) \sqrt{\frac{m}{V}} \left(U\left(1-\frac{mU}{V}\right) \frac{1+\sqrt{1-e^{-m}}}{\sqrt{1-\left(1-\frac{mU}{V}\right)}} + Ve^{-m} \frac{1+\sqrt{1-\left(1-\frac{mU}{V}\right)}}{\sqrt{1-e^{-m}}}\right) \\
        &= \frac{1}{2} \left(1+\frac{m}{V}\right) \sqrt{\frac{m}{V}} \left(U\left(1-\frac{mU}{V}\right) \frac{2-\frac{1}{2}e^{-m}}{\sqrt{\frac{mU}{V}}} + Ve^{-m} \frac{1+\sqrt{\frac{mU}{V}}}{1 - \frac{1}{2}e^{-m}}\right) \\
        &=\frac{1}{2} \left(1+\frac{m}{V}\right) \left(U\left(1-\frac{mU}{V}\right) \frac{2-\frac{1}{2}e^{-m}}{\sqrt{U}} + \sqrt{mV}e^{-m} \frac{1+\sqrt{\frac{mU}{V}}}{1 - \frac{1}{2}e^{-m}}\right) \\
        &=\frac{1}{2} \left(1+\frac{m}{V}\right) \left(\sqrt{U}\left(1-\frac{mU}{V}\right) \left(2-\frac{1}{2}e^{-m}\right) + \sqrt{m}e^{\frac{1}{2} \ln V-m} \frac{1+\sqrt{\frac{mU}{V}}}{1 - \frac{1}{2}e^{-m}}\right) \\
        &= \frac{1}{2} \cdot 1 \left(\sqrt{U} \cdot 1 \cdot 2 + \sqrt{m} e^{\frac{1}{2}\ln V-m} \frac{1}{1}\right) \\
        &= \sqrt{U} + \frac{1}{2} \sqrt{m} e^{\frac{1}{2}\ln V-m}.
\end{aligned}
\end{equation}
Note that $m$ is determined as a function of $U$ and $V$ by solving
\begin{equation}
m + \frac{1}{2} \ln m - \frac{1}{2}\ln V + \ln \left(4 \left(\sqrt{U} - 1\right)\right) = 0 \quad \text{for $V \gg 1$}.
\label{functionOfM}
\end{equation}
Hence,
\begin{equation}
\begin{aligned}
\frac{1}{2}\ln V - m &=  \frac{1}{2} \ln m + \ln K, \text{ where } K =  4 \left(\sqrt{U} - 1\right),
\end{aligned}
\end{equation}
and thus
\begin{equation}
\begin{aligned}
\Lambda &= \sqrt{U} + \frac{1}{2} \sqrt{m} e^{\frac{1}{2}\ln m + \ln K} = \sqrt{U} + \frac{K}{2}m.
\label{approLamuda2}
\end{aligned}
\end{equation}
From Eq.~\eqref{functionOfM}, we find $m$ is approximated by $\frac{1}{2} \ln V$ in the limit $V \to \infty$. Therefore, for $V \to \infty$, we have 
\begin{equation}
\ln \Lambda \simeq \ln \ln V.
\label{approLamuda3}
\end{equation}
So we conclude that
\begin{equation}
    \nu = \frac{ \ln U}{\ln \ln V}.
\end{equation}

\subsubsection{Percolating strength $C_\infty$}
The noncluster-defined $C_\infty$ is solved by:
\begin{equation}
\label{c_inf}
    \begin{cases}
        \left\{
        \begin{aligned}
        \text{seri}(x', t') &= \text{seri}(t, \text{para}(\text{seri}(x,\stackrel{a}{\overbrace{c,c,\dots,c}}),\text{seri}(y,\stackrel{U+V-1-a}{\overbrace{c,c,\dots,c}}))) \\
        \text{seri}(y', t') &= \text{seri}(t, \text{para}(\text{seri}(x,\stackrel{U-1-a}{\overbrace{c,c,\dots,c}}),\text{seri}(y,\stackrel{V+a}{\overbrace{c,c,\dots,c}}))) \\
        \text{seri}(x', y') &= \text{para}(\text{seri}(\stackrel{U}{\overbrace{c,c,\dots,c}}),\text{seri}(\stackrel{V}{\overbrace{c,c,\dots,c}})) 
        \end{aligned}
        \right.
        & \text{, where } a = 0, 1, 2, \dots, U-1;\\
\\
        \left\{
        \begin{aligned}
        \text{seri}(x', t') &= \text{seri}(t, \text{para}(\text{seri}(x,\stackrel{b}{\overbrace{c,c,\dots,c}}),\text{seri}(y,\stackrel{U+V-1-b}{\overbrace{c,c,\dots,c}}))) \\
        \text{seri}(y', t') &= \text{seri}(t, \text{para}(\text{seri}(x,\stackrel{V-1-b}{\overbrace{c,c,\dots,c}}),\text{seri}(y,\stackrel{U+b}{\overbrace{c,c,\dots,c}}))) \\
        \text{seri}(x', y') &= \text{para}(\text{seri}(\stackrel{U}{\overbrace{c,c,\dots,c}}),\text{seri}(\stackrel{V}{\overbrace{c,c,\dots,c}})) 
        \end{aligned}
        \right.
        & \text{, where } b = 0, 1, 2, \dots, V-1.\\
    \end{cases}
\end{equation}
When $n \to \infty$, $x'=y'=x=y=c^{\frac{1}{2}}$, the quantum series-parallel rules are
\begin{subequations}
\begin{align}
    \text{seri}(c_1,c_2) &= c_1 c_2,\\
    \text{para}(c_1,c_2) &= \left(1-\left(\dfrac{1}{2} \left(1+\sqrt{1-c_1^{2}}\right)\left(1+\sqrt{1-c_2^{2}}\right) -1\right)^2 \right) ^{\frac{1}{2}}.
\end{align}
\end{subequations}
Thus, $C_\infty$ is simplified to
\begin{equation}
\label{quantum u+v equations}
    \begin{small}
    \begin{cases}
        \left\{
        \begin{aligned}
        c^{\frac{1}{2}} t' &= t \left(1-\left(\dfrac{1}{2} \left(1+\sqrt{1-c^{2(a+\frac{1}{2})}}\right)\left(1+\sqrt{1-c^{2(U+V-\frac{1}{2}-a)}}\right) -1\right)^2 \right) ^{\frac{1}{2}} \\
        c^{\frac{1}{2}} t' &= t \left(1-\left(\dfrac{1}{2} \left(1+\sqrt{1-c^{2(U-\frac{1}{2}-a)}}\right)\left(1+\sqrt{1-c^{2(V+a+\frac{1}{2})}}\right) -1\right)^2 \right) ^{\frac{1}{2}} \\
        c^{\frac{1}{2}} c^{\frac{1}{2}} &= \left(1-\left(\dfrac{1}{2} \left(1+\sqrt{1-c^{2U}}\right)\left(1+\sqrt{1-c^{2V}}\right) -1\right)^2 \right) ^{\frac{1}{2}}
        \end{aligned}
        \right.
        & \text{, where } a = 0, 1, 2, \dots, U-1;\\
\\
        \left\{
        \begin{aligned}
        c^{\frac{1}{2}} t' &= t \left(1-\left(\dfrac{1}{2} \left(1+\sqrt{1-c^{2(b+\frac{1}{2})}}\right)\left(1+\sqrt{1-c^{2(U+V-\frac{1}{2}-b)}}\right) -1\right)^2 \right) ^{\frac{1}{2}} \\
        c^{\frac{1}{2}} t' &= t \left(1-\left(\dfrac{1}{2} \left(1+\sqrt{1-c^{2(V-\frac{1}{2}-b)}}\right)\left(1+\sqrt{1-c^{2(U+b+\frac{1}{2})}}\right) -1\right)^2 \right) ^{\frac{1}{2}} \\
        c^{\frac{1}{2}} c^{\frac{1}{2}} &= \left(1-\left(\dfrac{1}{2} \left(1+\sqrt{1-c^{2U}}\right)\left(1+\sqrt{1-c^{2V}}\right) -1\right)^2 \right) ^{\frac{1}{2}}
        \end{aligned}
        \right.
        & \text{, where } b = 0, 1, 2, \dots, V-1.\\
    \end{cases}
    \end{small}
\end{equation}
By Eq.~\eqref{quantum u+v equations}, the average value satisfies
\begin{equation}
    \begin{scriptsize}
    \begin{aligned}
        \dfrac{t'}{t} &= \dfrac{\sum_{a=0}^{U-1} \left( \left(1-\left(\dfrac{1}{2} \left(1+\sqrt{1-c^{2(a+\frac{1}{2})}}\right)\left(1+\sqrt{1-c^{2(U+V-\frac{1}{2}-a)}}\right) -1\right)^2 \right)
        \left(1-\left(\dfrac{1}{2} \left(1+\sqrt{1-c^{2(U-\frac{1}{2}-a)}}\right)\left(1+\sqrt{1-c^{2(V+a+\frac{1}{2})}}\right) -1\right)^2 \right) \right) ^{\frac{1}{4}} } 
        {\left(U+V \right) \left(1-\left(\dfrac{1}{2} \left(1+\sqrt{1-c^{2U}}\right)\left(1+\sqrt{1-c^{2V}}\right) -1\right)^2 \right) ^{\frac{1}{4}}} \\
        &+ \dfrac{\sum_{b=0}^{V-1} \left( \left(1-\left(\dfrac{1}{2} \left(1+\sqrt{1-c^{2(b+\frac{1}{2})}}\right)\left(1+\sqrt{1-c^{2(U+V-\frac{1}{2}-b)}}\right) -1\right)^2 \right) 
        \left(1-\left(\dfrac{1}{2} \left(1+\sqrt{1-c^{2(V-\frac{1}{2}-b)}}\right)\left(1+\sqrt{1-c^{2(U+b+\frac{1}{2})}}\right) -1\right)^2 \right) \right) ^{\frac{1}{4}}}
        {\left(U+V \right) \left(1-\left(\dfrac{1}{2} \left(1+\sqrt{1-c^{2U}}\right)\left(1+\sqrt{1-c^{2V}}\right) -1\right)^2 \right) ^{\frac{1}{4}}}.
    \end{aligned}
    \end{scriptsize}
\end{equation}
We denote 
\begin{equation}
    \begin{scriptsize}
    \begin{aligned}
        f_3 &= \sum_{a=0}^{U-1} \left( \left(1-\left(\dfrac{1}{2} \left(1+\sqrt{1-c^{2(a+\frac{1}{2})}}\right)\left(1+\sqrt{1-c^{2(U+V-\frac{1}{2}-a)}}\right) -1\right)^2 \right)
        \left(1-\left(\dfrac{1}{2} \left(1+\sqrt{1-c^{2(U-\frac{1}{2}-a)}}\right)\left(1+\sqrt{1-c^{2(V+a+\frac{1}{2})}}\right) -1\right)^2 \right) \right) ^{\frac{1}{4}} \\
        &\stackrel{V \to \infty} \simeq \sum_{a=0}^{U-1} 
        \left( \left(1-\left(\dfrac{1}{2} \left(1+O(\sqrt{\frac{\ln{V}}{V}}) \right)\left(1+\sqrt{1-e^{-m}}\right) -1\right)^2 \right)
        \left(1-\left(\dfrac{1}{2} \left(1+O(\sqrt{\frac{\ln{V}}{V}}) \right)\left(1+\sqrt{1-e^{-m}}\right) -1\right)^2 \right) \right) ^{\frac{1}{4}}\\
        &\stackrel{V \to \infty} \simeq U \left(1-O\left(\sqrt{\frac{\ln{V}}{V}}\right) - O\left(\frac{1}{\sqrt{V}} \right) \right) ^{\frac{1}{2}} \\
        &\simeq U + O\left(\sqrt{\frac{\ln{V}}{V}}\right),
    \end{aligned}
    \end{scriptsize}
\end{equation}
and
\begin{equation}
    \begin{scriptsize}
    \begin{aligned}
        f_4 &= \sum_{b=0}^{V-1} \left( \left(1-\left(\dfrac{1}{2} \left(1+\sqrt{1-c^{2(b+\frac{1}{2})}}\right)\left(1+\sqrt{1-c^{2(U+V-\frac{1}{2}-b)}}\right) -1\right)^2 \right) 
        \left(1-\left(\dfrac{1}{2} \left(1+\sqrt{1-c^{2(V-\frac{1}{2}-b)}}\right)\left(1+\sqrt{1-c^{2(U+b+\frac{1}{2})}}\right) -1\right)^2 \right) \right) ^{\frac{1}{4}} \\
        &\stackrel{V \to \infty} \simeq \int_{(c^2)^{\frac{1}{2}}}^{(c^2)^{V-\frac{1}{2}}} 
        \left(1-\left(\dfrac{1}{2} \left(1+\sqrt{1-c^{2(b+\frac{1}{2})}}\right)\left(1+\sqrt{1-c^{2(U+V-\frac{1}{2}-b)}}\right) -1\right)^2 \right)^{\frac{1}{4}}\\ 
        &\left(1-\left(\dfrac{1}{2} \left(1+\sqrt{1-c^{2(V-\frac{1}{2}-b)}}\right)\left(1+\sqrt{1-c^{2(U+b+\frac{1}{2})}}\right) -1\right)^2 \right)^{\frac{1}{4}} \frac{d(c^2)^{b + \frac{1}{2}}}{(c^2)^{b+\frac{1}{2}} \ln{(c^2)}} \\
        &\stackrel{V \to \infty} \simeq \int_{(c^2)^{\frac{1}{2}}}^{(c^2)^{V-\frac{1}{2}}} \frac{-V}{m} \frac{1}{x}
        \left( \left(1-\left(\dfrac{1}{2} \left(1+\sqrt{1-x}\right) \left(1+\sqrt{1-e^{-m} x^{-1}}\right) -1\right)^2 \right) 
        \left(1-\left(\dfrac{1}{2} \left(1+\sqrt{1-e^{-m} x^{-1}}\right) \left(1+\sqrt{1-x}\right) -1\right)^2 \right) \right) ^{\frac{1}{4}} \, dx \\
        &\simeq \int_{1}^{e^{-m}} \frac{1}{x} \frac{-V}{m} x^{\frac{1}{2}} \, dx \\
        &\simeq \frac{2V}{m} \left(1-e^{-\frac{m}{2}} \right) \\
        &\simeq \frac{4V}{\ln{V}} + O \left(\dfrac{V \ln{\ln{V}}}{(\ln{V})^2} \right).
    \end{aligned}
    \end{scriptsize}
\end{equation}
Therefore, 
\begin{equation}
    \begin{aligned}
        \dfrac{C'_\infty}{C_\infty} &= \dfrac{t'}{t} \\
        &= \dfrac{f_3 + f_4}{\left(U+V \right) \left(1-\left(\dfrac{1}{2} \left(1+\sqrt{1-c^{2U}}\right)\left(1+\sqrt{1-c^{2V}}\right) -1\right)^2 \right) ^{\frac{1}{4}} } \\
        &\simeq \dfrac{\frac{4V}{\ln{V}} + O \left( \frac{V \ln{\ln{V}}} {(\ln{V})^2} \right)} {V \left( 1+ O\left( \sqrt{\frac{\ln{V}}{V}}\right) \right)} \\
        &\simeq \dfrac{4}{\ln{V}} + O\left( \dfrac{\ln{\ln{V}}}{(\ln{V})^2}\right).
    \end{aligned}
\end{equation}

\subsubsection{Critical exponents $d_f$ and $\beta$}
Similar to the classical case, 
\begin{equation}
    \begin{aligned}
        d_{f} &\simeq \frac{\ln{V}}{\ln{U}} \left(1-\dfrac{\ln{\ln{V}} - 2\ln2}{\ln{V}} + O\left( \dfrac{\ln{\ln{V}}}{(\ln{V})^2}\right) \right)\\
        &\simeq \dfrac{\ln{V}}{\ln{U}} - \dfrac{\ln{\ln{V}}}{\ln{U}} + O\left( \dfrac{\ln{\ln{V}}}{\ln{V}}\right).
    \end{aligned}
\end{equation}
Assuming the hyperscaling relation $\beta = \nu (d-d_{f})$, we derive
\begin{equation}
    \begin{aligned}
        \beta &\simeq \left( \frac{\ln U}{\ln{\ln{\sqrt{V}}}} + O\left(\frac{\ln{\ln{\ln{V}}}}{(\ln{\ln{V}})^2}\right) \right) \left( \dfrac{\ln{\ln{V}}}{\ln{U}} + O\left( \dfrac{\ln{\ln{V}}}{\ln{V}}\right) \right)\\
        &\simeq 1 + O\left(\dfrac{\ln{\ln{\ln{V}}}}{\ln{\ln{V}}}\right).
    \end{aligned}
\end{equation}

\subsection{\, Comparison with numerical results}

\begin{figure}[ht!]
\centering
\includegraphics[width=\linewidth]{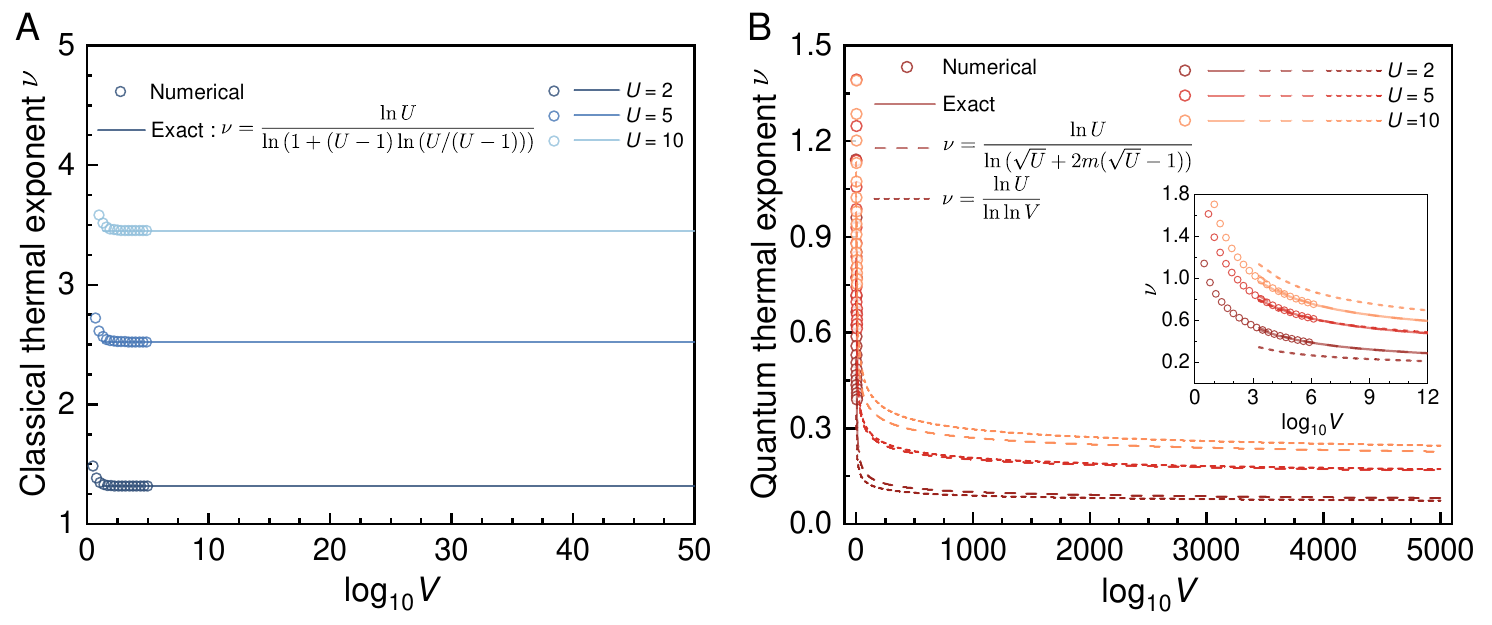}
\caption{\label{nu solutions} \textbf{The critical exponent $\nu$ as a function of $V$ when the shortest path $U=2,5,10$.} 
(\textbf{A})~Classical percolation. Open blue circles are simulation values using finite-size analysis. The colors from darker blue to lighter blue are for $U=2,5,10$ respectively. Solid line is derived theoretical solution, which is a constant equal to $\ln{U}/\ln{(1+\ln{U})}$ for $V\to \infty$ and is larger when U increases. 
(\textbf{B})~For quantum percolation, when $U$ is fixed, $\nu$ exhibits a different asymptotic behavior. The lines from top to bottom demonstrates decreasing degree of approximation [Eqs.~\eqref{approLamuda1},~\eqref{approLamuda2}, and~\eqref{approLamuda3}], respectively for $\Lambda$ in the equation $\nu = \ln{U}/\ln{\Lambda}\big|_{c=c_\text{th}}$. Open red circles are simulation values using finite-size analysis. The solid light red line is derived theoretical solution. The colors from darker orange to lighter orange are for $U=2,5,10$ respectively. The quantum $\nu$ decreases as $V$ increases, and will eventually reach $\nu = 0$ for any constant $U$. Even though $\nu \to 0$ for $V \to \infty$, the nonzero values of $\nu$ observed for $V$ finite $(V\approx 10^{5000})$ might be of physical relevance for a QN.}
\hfill\hfill
\end{figure}
For ($U,V$) flowers with different $U$ and $V$, these critical exponents have very different asymptotic behaviors. We generalize above results to the asymptotic limits of constant $U$ and $V \rightarrow \infty$ and find that the classical exponent $\nu$ does not change. For example, $U=2$, given the expression of the classical exponent $\nu$, it must be greater than the lower bound $\nu\approx 1.316$ (Fig.~\ref{nu solutions}A) when $V$ increase. By comparison, in the quantum case the exponent $\nu$ decreases with $V$ and will eventually reach the value of 0 for $V \rightarrow \infty$ (Fig.~\ref{nu solutions}B). These results demonstrate that the classical exponent $\nu$ is independent on $V$ while the quantum exponent $\nu$ changes with both $U$ and $V$. 
In addition, for larger constant $U$, the limit value of the classical exponent $\nu$ also increases, implying that $U$ plays a decisive role in the classical case. While for the quantum case, the $\nu$ for $U=2,5,10$ are closer compared with the classical case. But the distinctions still exist in the speed of going to zero due to the weighting factor related to the shortest path $U$ in quantum percolation.

\section{Analysis on real-world Internet network}
\begin{figure}[ht!]
    \centering
    \includegraphics[width=\linewidth]{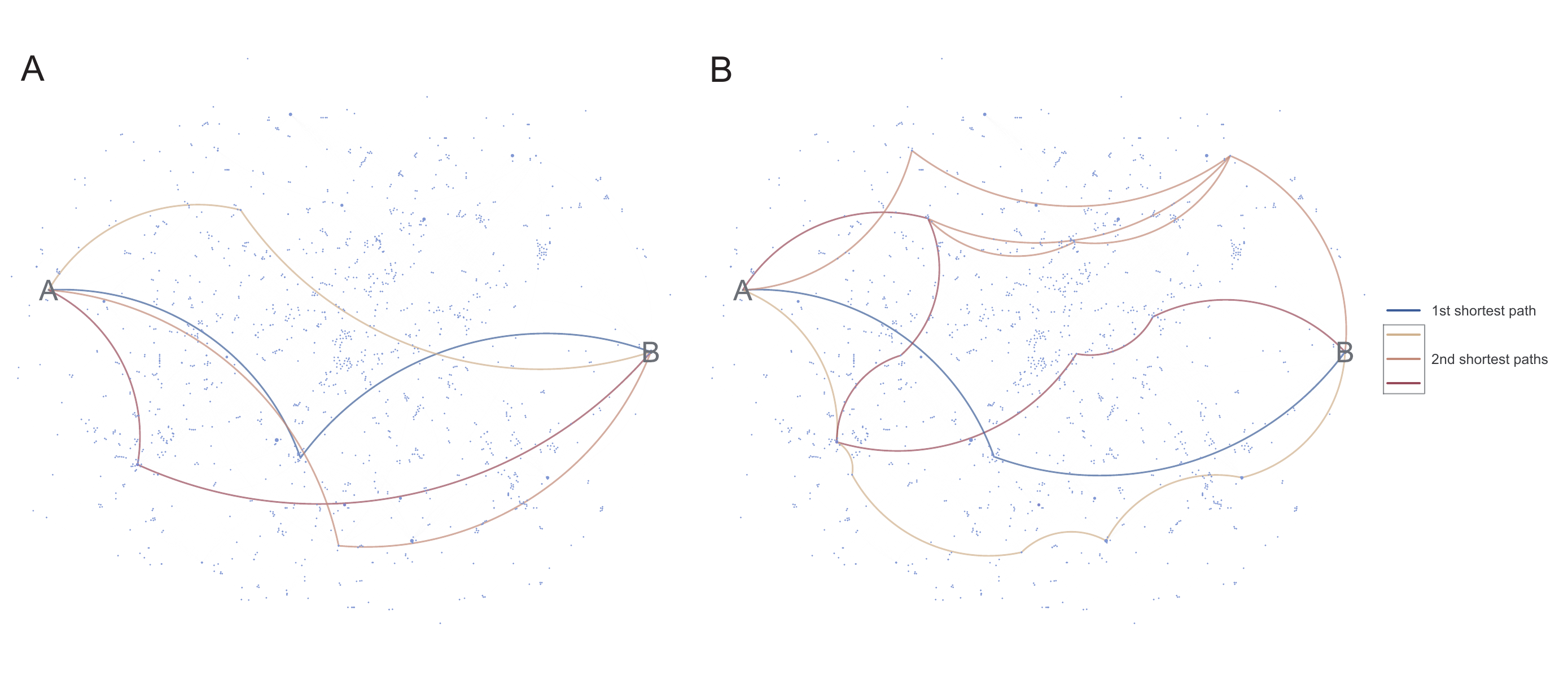}
    \caption{\textbf{Illustration of the Internet including a focused subgraph of the Internet, centered around two hubs \textit{A} and \textit{B}.}
    The subgraphs highlight multiple hierarchical and intersecting paths of varying lengths, ranging from the shortest to longer routes. (\textbf{A})~The initial subgraph $g_1$ ($q=1$). (\textbf{B})~The subgraph $g_q$ with $q$-times longer paths between the same hubs ($q=3$).
    \hfill\hfill}
    \label{fig_internet}
\end{figure}

We analyzed the effect of longer paths in concurrence percolation compared to classical percolation on a real autonomous-system-level Internet network (Fig.~\ref{fig_internet}). The detouring process required that the path lengths of level $k$ in the network should grow by a factor of $q^k$. We focused on two distinct levels of path lengths, corresponding to path numbers of $(1, 3)$, respectively; thus the growth multiples were $(1, q)$.

The detouring process goes as follows: First, an initial subgraph $g_1$ ($q=1$) was constructed, composed of four shortest nonoverlapping paths between two randomly selected nodes. On this basis, for $q>1$, the first-level shortest path was kept unchanged, while the three second-level paths were replaced by $q$-times longer paths between the same nodes, producing a new subgraph $g_q$. All paths must have no overlaps.
During the process, the selection of nodes was flexible, with the only constraint that all the subgraphs with nonoverlapping paths were able be found between them. Under this constraint, we selected $10$ random pairs of nodes with large enough node degrees, which should be no less than $7$. At the same time, since all the paths of $g_q$ were only length-fixed, it was possible to find more than one $g_q$ from the original real network. We randomly selected $20$ $g_q$ for each pair of nodes. As a result, for each $q$, the simulation involved $10$ different $g_1$ and $200$ different $g_q$ in total.

We located the finite-size percolation thresholds of all $g_q$ under different path lengths at $P_\text{sc}=0.99$ (in classical case) or $C_\text{sc}=0.99$ (in quantum case), where $q=2,3,\dots,8$. For each $q$, the thresholds were averaged and substituted into $A_x(q)$ [Eq.~\eqref{Axq} in the main text], yielding the final results (Fig.~\ref{fig_real net}C in the main text). 

\end{document}